\documentclass[a4paper,11pt]{article}
\usepackage{jheppub}
\usepackage{graphicx}
\usepackage{amsmath}
\usepackage{amsfonts}
\usepackage{amssymb}
\usepackage{slashed}
\usepackage[colorlinks=true,linkcolor=blue]{hyperref}
\usepackage{bm}
\usepackage{empheq}
\usepackage[super]{nth}
\def\sss{\scriptscriptstyle}
\def\ah{a_{\sss{H}}}
\def\tA{\theta_{\!\sss{A}}}
\def\tAstr{\theta^{\,\mathrm{str}}_{\!\sss{A}}}
\def\tcut{t_{\mathrm{cut}}}
\def\rmin{r_{\mathrm{min}}}

\def\half{\frac{1}{2}}

\def\LL{\mathcal{L}}
\def\be{\begin{equation}}
\def\ee{\end{equation}}
\def\bea{\begin{eqnarray}}
\def\eea{\end{eqnarray}}

\def\Eq#1{Eq.~\eqref{#1}}
\def\Re{\,\mathrm{Re}\,}

\def\nn{\nonumber}
\def\nax{n_{\mathrm{ax}}}
\def\atan{\mathrm{atan}}

\title{Axion String Dynamics I:  2+1D}
\author[1]{Leesa M.\ Fleury,}
\author[2]{Guy D.\ Moore}

\affiliation[1]{University of British Columbia, Department of Physics \& Astronomy,\\6224 Agricultural Road, Vancouver BC V6T 1Z1, Canada}
\affiliation[2]{Institut f\"ur Kernphysik, Technische Universit\"at Darmstadt\\
Schlossgartenstra{\ss}e 2, D-64289 Darmstadt, Germany}
\emailAdd{lfleury@phas.ubc.ca}
\emailAdd{guy.moore@physik.tu-darmstadt.de}

\abstract{
If the axion exists and if the initial axion field value is
uncorrelated at causally disconnected points, then it should be
possible to predict the efficiency of cosmological axion production,
relating the axionic dark matter density to the axion mass.  The main
obstacle to making this prediction is correctly treating the axion
string cores.  We develop a new algorithm for treating the axionic
string cores correctly in 2+1 dimensions.  When the axionic string
cores are given their full physical string tension, axion production is about twice as
efficient as in previous simulations.  We argue that the string
network in 2+1 dimensions should behave very differently than in 3+1
dimensions, so this result cannot be simply carried over to the
physical case.  We outline how to extend our method to 3+1D axion
string dynamics.
}

\begin{document}
\maketitle
\section{Introduction}
\label{sec:intro}

The axion
\cite{Peccei:1977hh,Peccei:1977ur,Weinberg:1977ma,Wilczek:1977pj}
is a hypothetical particle, the angular excitation of a
proposed ``Peccei-Quinn'' (PQ) field $\varphi$ that would solve the strong
CP problem \cite{tHooft:1976,Jackiw:1976pf,Callan:1979bg} while
providing a viable dark matter candidate
\cite{Preskill:1982cy,Abbott:1982af,Dine:1982ah}.
If PQ symmetry is restored in the early Universe (either during or
after inflation), it should be possible to make a definite prediction
for the axion dark matter density which would be produced
cosmologically.  However, the axion production is complicated by
topological structures (axionic strings) which appear in the axionic
field \cite{Davis:1986xc}, and so far their dynamics have not been
reliably simulated.
Thus we currently lack a quantitative determination of the
efficiency of axion production in this scenario, and can therefore not
yet fix the relation between the axion dark matter density and the
axion mass.  This is unfortunate.  If we could
fix this relation, it would make the axion dark-matter scenario
predictive and help axion search experiments
\cite{Sikivie:1983ip,Bradley:2003kg,Asztalos:2009yp} know in what
frequency bandwidth to look.

The goal of this paper is to make some progress on developing the
tools to study axion production in the presence of axionic strings.
We will not present a complete methodology or reliable results, but we
lay the groundwork for getting there by presenting
an interesting algorithmic advance.  We
start by reviewing the relevant physics of the axion field, and
clarifying the main problem.  The axion field is a complex
scalar $\varphi$ which spontaneously breaks a $U(1)$ (phase)
symmetry.  The Lagrangian density is
\be
\label{LLbare}
-\LL_{\varphi_1} = \partial_\mu \varphi^* \partial^\mu \varphi
+ \frac{\lambda}{8} \left(2 \varphi^* \varphi - f_a^2 \right)^2 \,,
\ee
invariant under $\varphi \to \varphi e^{i\theta}$,
plus a small explicit breaking term, arising from QCD axions:
\be
\label{LLbreak}
-\LL = -\LL_{\varphi_1} +
\chi(T) \left[ 1 - \cos\left(\mathrm{arg}\:\varphi\right) \right],
\ee
with $f_a \sim 10^{11}$GeV the axion decay constant (the vacuum value
for the $\varphi$ field) and
$\chi(T)$ the temperature-dependent topological susceptibility of
QCD.  The decay constant $f_a$ is a model parameter, and $\chi(T)$ is
a calculable quantity in QCD, which is currently not well determined
at high temperature \cite{Borsanyi:2015cka,diCortona:2015ldu}.
We will not address the problem of finding $\chi(T)$ here.  Instead we
focus on the topological consequences of a spontaneously broken global
$U(1)$ symmetry, which is also very weakly explicitly broken.

\begin{figure}[htb]
\centerline{  \epsfxsize=0.45\textwidth\epsfbox{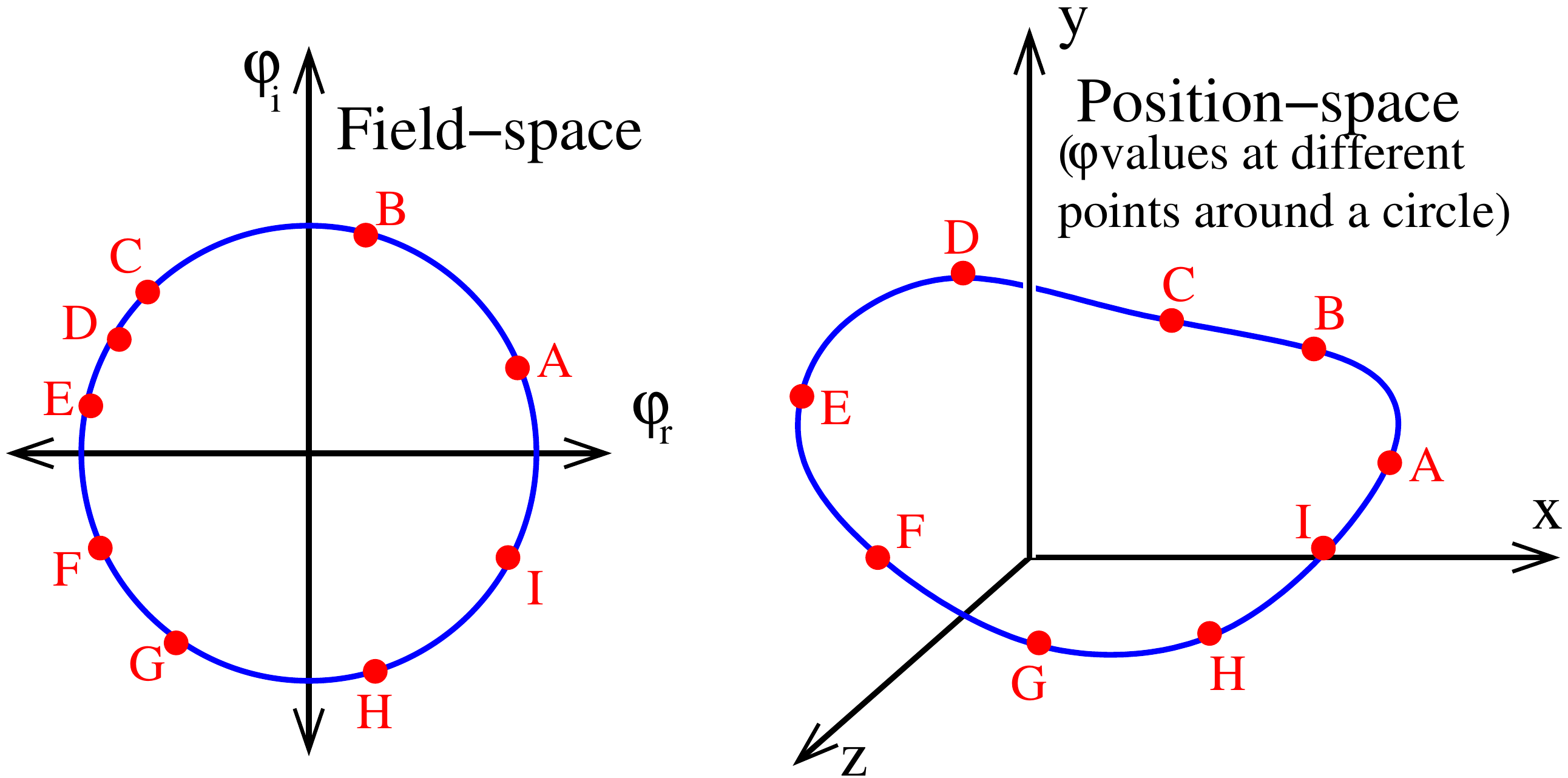} \hfill
  \epsfxsize=0.48\textwidth\epsfbox{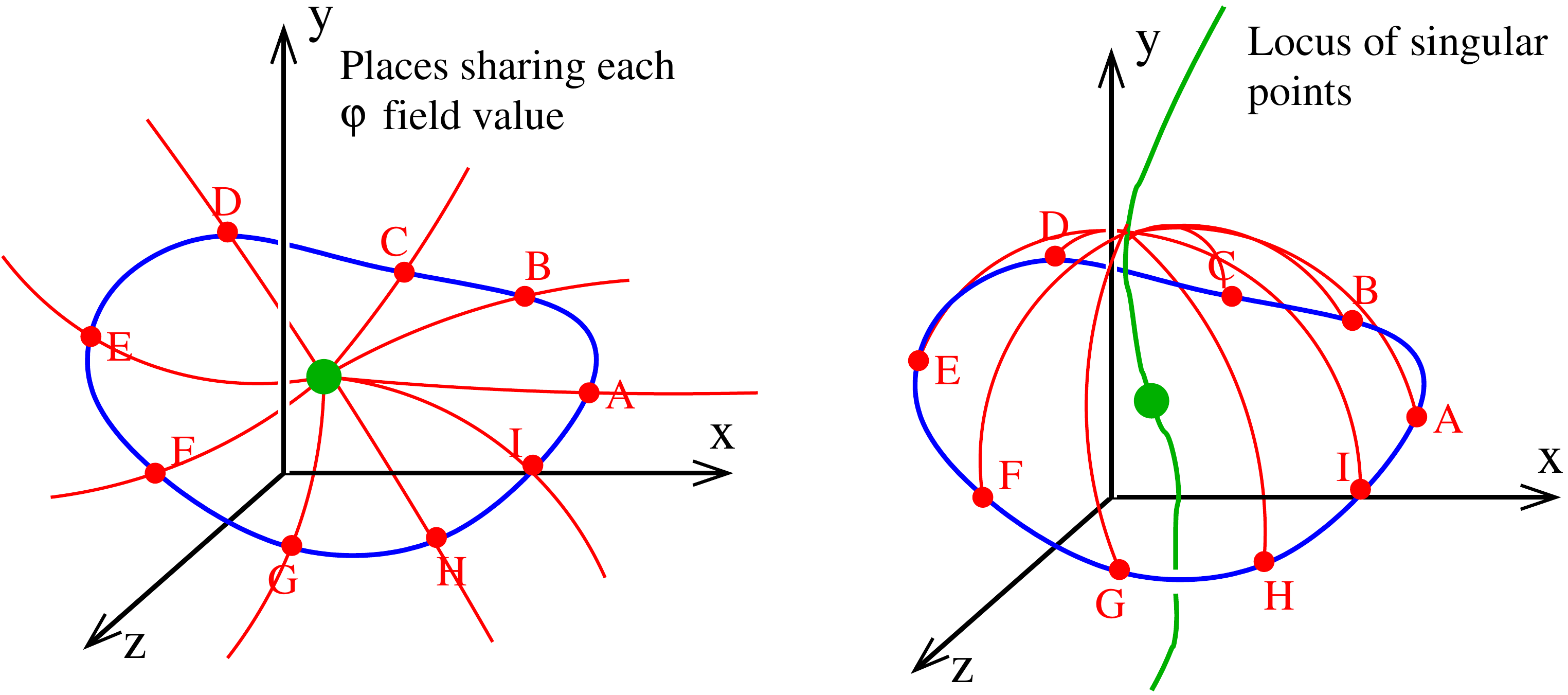}}
  \caption{\label{string_illust}
    Illustration of the topological origin of strings.  A series of
    points along the minimum of the potential in field-space
    (leftmost) may be mapped onto by a series of points along a
    loop in position-space (next-to-left).  On a surface
    bounded by the coordinate-space loop, the set of points which
    map onto $A$, $B$, \textsl{etc}.\ each form a curve
    (third-from-left) which must meet at a singularity in $\tA$.  By
    considering a family of surfaces, there must be a locus of such
    singular points forming a string (rightmost).}
\end{figure}

The $\varphi$ field has heavy radial excitations with mass-squared
$m_s^2 \equiv \lambda f_a^2$ and nearly-massless angular excitations with
mass-squared $m_a^2 = \chi(T)/f_a^2$.
At temperatures $T \ll f_a$ and length scales $r\gg f_a^{-1}$ we can
treat $\varphi$ as a classical field and treat the radial excitations
as heavy, so the field is almost-everywhere on the
``vacuum manifold'' $\varphi^* \varphi = 2f_a^2$.  Then we can write
$\sqrt{2}\:\varphi = f_a e^{i\tA}$, with $\tA=\mathrm{arg}\:\varphi$
the ``axion angle''; the potential energy is only very weakly
dependent on the value of $\tA$ through the symmetry breaking term,
which at high temperatures (roughly $T>1.5$ GeV) can be ignored.  The
angle $\tA$ is only defined modulo $2\pi$.
It is possible for the field to leave the vacuum
manifold along a linelike defect, with $\tA$ varying by $2\pi$ around
any loop which circles the linelike defect in a positive sense
(see Fig.~\ref{string_illust}).

\noindent
Such a defect -- essentially a vortex in the $\varphi$ field -- is
called an axionic
cosmic string, and it is topologically stable; no local changes to the
value of $\varphi$ can cause it to disappear.  If PQ symmetry is
restored in the early Universe, then $\tA$ starts out uncorrelated at
widely separated points and will generically begin with a dense
network of these strings (the Kibble mechanism for string production
\cite{Kibble}).  The strings evolve, straightening out, chopping off
loops, and otherwise reducing their density, arriving at a scaling
solution \cite{Albrecht:1984xv} where the length of string per unit 
volume scales with time t as $t^{-2}$ up to logarithmic corrections
(which we will discuss).

Let us analyze the structure of a string in a little more detail.
Consider a straight string along the $z$ axis; in polar $(z,r,\phi)$
coordinates the string equations of motion are solved by
$\sqrt{2} \varphi = v(r) f_a e^{i\phi}$, with $v(r) \simeq 1$ for all
$r^2 \gg 1/(\lambda f_a^2)$; so $\tA = \phi$ (up to a constant which
we can remove by our choice of $x$-axis).  The string's energy is
dominated by the gradient energy due to the space variation of $\tA$:
\bea
\label{Tension}
T &=& \frac{\mbox{Energy}}{\mbox{length}}
= \int r\,dr\,d\phi \left( V(\varphi^* \varphi) + \half \nabla \varphi^*
\nabla \varphi \right)
 \\
\label{kappa}
 & \simeq & \pi \int r\, dr \left( \frac{\partial_\phi \varphi^*}{r}\;
\frac{\partial_\phi \varphi}{r} \right) \simeq
\pi \int^{H^{-1}}_{1/m_s} r \, dr \; \frac{f_a^2}{r^2}
= \pi f_a^2 \ln(m_s /H) \equiv \pi f_a^2 \kappa \,,
\eea
where the integral over $r$ is cut off at small $r$ by the scale where
$v(r) \neq 1$ (the string core), and at large distances by the scale
where the string is not alone in the Universe but its field is
modified by other strings or effects, which we took to be the Hubble
scale $H^{-1}$.  We define
$\kappa = \ln(m_s/H) = \ln(f_a \sqrt{\lambda}/H)$ as
the ratio of these scales; typically $f_a \sim 10^{11}$ GeV, while at
the relevant temperature range $T \sim 1$ GeV the Hubble scale is
$H\sim 10^{-18}$ GeV, so $\kappa \sim 70$ (unless $\lambda$ is orders
of magnitude smaller than 1).

This logarithm, $\kappa$, controls several aspects of the strings'
dynamics.  It controls the string tension, as we just saw.  More
relevant, while the string tension is $\pi \kappa f_a^2$, the
string's interactions with the long-range $\varphi$ field scale as
$f_a^2$.  Therefore the string's long-range interactions become less
important, relative to the string evolution under tension, as $\kappa$
gets larger.  The long-range interactions are responsible for energy
radiation from the strings, as well as for long-range, often
attractive, interactions between strings.  Since these effects tend to
deplete and straighten out the string network, the large-$\kappa$ theory
will have denser, cuspier strings.  Indeed, in the large $\kappa$
limit the string behavior should go over to that of local (Nambu-Goto)
strings \cite{Dabholkar:1989ju}, which are known to have far denser
string networks than axionic networks with $\kappa \sim 6$.

In our first paper on this subject \cite{axion1}, we simulated string
networks, with and without the potential-tilting term of \Eq{LLbreak},
in ``field-only'' classical lattice simulations where the string cores
arose naturally from treating both radial and angular components of
the $\varphi$ field.%
\footnote{%
  There have been other field-only classical field theory simulations
  \cite{Yamaguchi:1998gx,%
    Yamaguchi:1999yp,%
    Hiramatsu:2010yu,%
    Hiramatsu:2012gg}, with similar results but with less dynamic
  range of $\kappa$ values explored.}
Therefore, the scales $m_s$ and $H$ both had to
be resolved on the lattice, which restricts the ratio to be less than
the number of points across the lattice $N$.  Realistically
$N\leq 2^{11}$ in 3D simulations and $N<2^{16}$ in 2D, which limited
us to studying the range $\kappa \leq 6$ in 3D and 8 in 2D.
We followed the string evolution through the regime where the
potential tilt, \Eq{LLbreak}, becomes important and even dominant.
Domain walls develop and destroy the string network \cite{Vilenkin:1982ks},
leaving behind axionic fluctuations whose density we seek to
determine.  Over the $\kappa$ range we could observe, we saw
clearly that the density of strings depends strongly on $\kappa$.
We also found that the behavior of the
system in 3+1 dimensions, in terms of string density, energy density,
and final axion number produced,
is surprisingly similar to that in 2+1D, by
which we mean 3+1D but with all fields constant along the $z$
direction.  Finally, we found
that the axion production rate was a surprisingly weak function of
$\kappa$.  Indeed, over the range we studied, axion production
actually decreased slightly as we raised $\kappa$.

Unfortunately, the physical value of $\kappa \sim 70$ is an order of
magnitude larger than we could achieve with field-only methods.  The
density of strings is presumably much larger for $\kappa=70$, as
predicted by one-scale models \cite{Martins:2000cs}.
And we know the string tension scales linearly with $\kappa$, as in
\Eq{kappa}.  It seems reasonable to expect that, as these denser
networks of higher-energy strings decay into axionic excitations,
the final axion number density will be higher than in the
small-$\kappa$ simulations (though we see no sign of this for the
$\kappa$ range we have been able to study).  But to verify this
suspicion, and really learn the axion production efficiency, we need
to simulate axion production
using this larger $\kappa$ value.  This paper will show how to
do this in full detail in 2+1 dimensions, and will argue for how to
extend the methodology to 3+1 dimensions.  
The basic idea of the simulations is to implement the angular component
$\tA$ of the $\varphi$ field on the lattice, while implementing
string cores as additional explicit objects (not restricted to the lattice) 
which interact appropriately with the lattice $\tA$ field.
Conceptually this is similar to the work of Dabholkar and
Quashnock \cite{Dabholkar:1989ju}; but their work was analytical and
did not present an algorithmic implementation for the lattice.
In the next section, we explain the method for the 2+1 dimensional
problem, by taking advantage of a dual electromagnetic description.
Section \ref{latt} presents the details of the lattice implementation.
In Section \ref{results} we explore the numerical results.  These lead
us to expect that the physics of 2+1D and 3+1D become ever more
different as the string cores get higher-tension.  Therefore, while
our results are suggestive, we cannot interpret them too literally for
the interesting 3+1D case.  We close by describing how our algorithm
can be extended to 3+1 dimensions.

\section{The dual electromagnetic picture}
\label{dualEM}

The basic idea of our method is the following.  At large distances
$r\sim H^{-1}$ or $r \sim m_a^{-1}$ the $\tA$ field displays
complicated dynamics which we need to solve nonperturbatively, via
lattice simulations.  The field also contains topological defects.
The cores of the defects involve very short scales, as we have already
emphasized.  But the physics of these string cores is actually very
simple, and we understand it analytically.  On short scales the string
is very straight, and in its local rest-frame the $\tA$ field varies
around the string with the angular $\phi$-coordinate, up to
corrections subleading in $m_a r$ or $Hr$.

It is a waste of effort to try to simulate the microscopic behavior of
the string core by solving the field equations of motion.  Instead we
should ``cut it out'' from our lattice and ``sew back in'' an explicit
object which reproduces the string core's behavior.  The main
challenge is to incorporate correctly the physics of how the string
influences $\tA$ in its environment, and how the environment
influences the string evolution.  In this section we will show how to
do that -- for the 2+1 dimensional theory.

\subsection{2+1D string defects as electromagnetic charges}

Consider the axion model in 2 space dimensions.  For ease of
presentation we will explain the approach in flat, non-expanding
space.  It is straightforward to re-introduce Hubble drag and to work
in terms of conformal time%
\footnote{%
  Axion number is set around $T\sim 1$GeV, when the universe is
  radiation dominated and the number of relativistic degrees of
  freedom $g_*$ is nearly constant.  The Hubble parameter is
  $H=\partial_t \ah /\ah=1/2t_{\mathrm{true}}$.  Conformal time
  $t_{\mathrm{conf}}$, henceforward just $t$, is
  $dt=dt_{\mathrm{true}}/\ah$.  The temperature scales as
  $T\propto t^{-1}$ and the metric scales as
  $g_{\mu\nu} \propto t^2 \eta_{\mu\nu}$.  Therefore mass scales in
  these units grow with an extra $t$ factor relative to the physical
  mass.  Our sign conventions are
  that $\eta_{\mu\nu} = \mathrm{Diag}\:[-1,+1,+1,+1]$ and
  $\epsilon_{0123}=1=-\epsilon^{0123}$. }.
Except within a tiny distance $r\sim 1/m_s$ of a string core,
the axion field is
determined by $\tA$ alone; ignoring for now the symmetry-breaking
potential term, its Lagrangian and equation of motion are
\be
  -  \LL_{\tA} = \frac{f_a^2}{2} \partial_\mu \tA \partial^\mu \tA \,, \qquad
  \partial_\mu \partial^\mu \tA = 0 \,.
\label{LtA}
\ee
A ``string'' defect in 2+1 dimensions is a point (monopole) defect,
with $\nabla \tA$ diminishing as $1/r$ radiating out from the defect.
This is the same falloff as the electric field of a charge in 2+1
dimensions.  The potential between two strings also has the same
$-\frac{q_1 q_2}{2\pi} \ln(r)$ form as in 2+1D electromagnetism.
Indeed, there is actually a perfect analogy between the
axion field and electromagnetism \cite{Hecht:1990mv}.  If we define
\be
  F_{\mu\nu} = - f_a \epsilon_{\mu\nu\alpha} \partial^\alpha \tA
\label{defF}
\ee
with $E_i = F_{i0} = F^{0i}$ and $B = F_{12}$, so in components%
\footnote{Recall that in 2+1 dimensions, the magnetic field is a
  pseudoscalar (corresponding to $B_z$ in 3+1D).}
\be
\label{EB}
E_i = f_a \epsilon_{ij} \partial_j \tA \,, \qquad
B = f_a \partial_t \tA \,,
\ee
then away from string cores these fields obey Maxwell's equations,
\bea
\label{Ampere}
\partial_\mu F^{\nu\mu} &=& -f_a \epsilon^{\nu\mu\alpha} \partial_\mu
\partial_\alpha \tA = 0 \,, \\
\epsilon_{\mu\nu\alpha} \partial^\alpha F^{\mu\nu}
& = & - f_a \epsilon_{\mu\nu\alpha} \epsilon^{\mu\nu \beta}\partial^\alpha
\partial_\beta \tA = 2 f_a \partial_\alpha \partial^\alpha \tA = 0 \,,
\label{Faraday}
\eea
where the first holds by antisymmetry and the second requires the
$\tA$ equation of motion.

\begin{figure}[htb]
  \centerline{\epsfxsize=0.7\textwidth\epsfbox{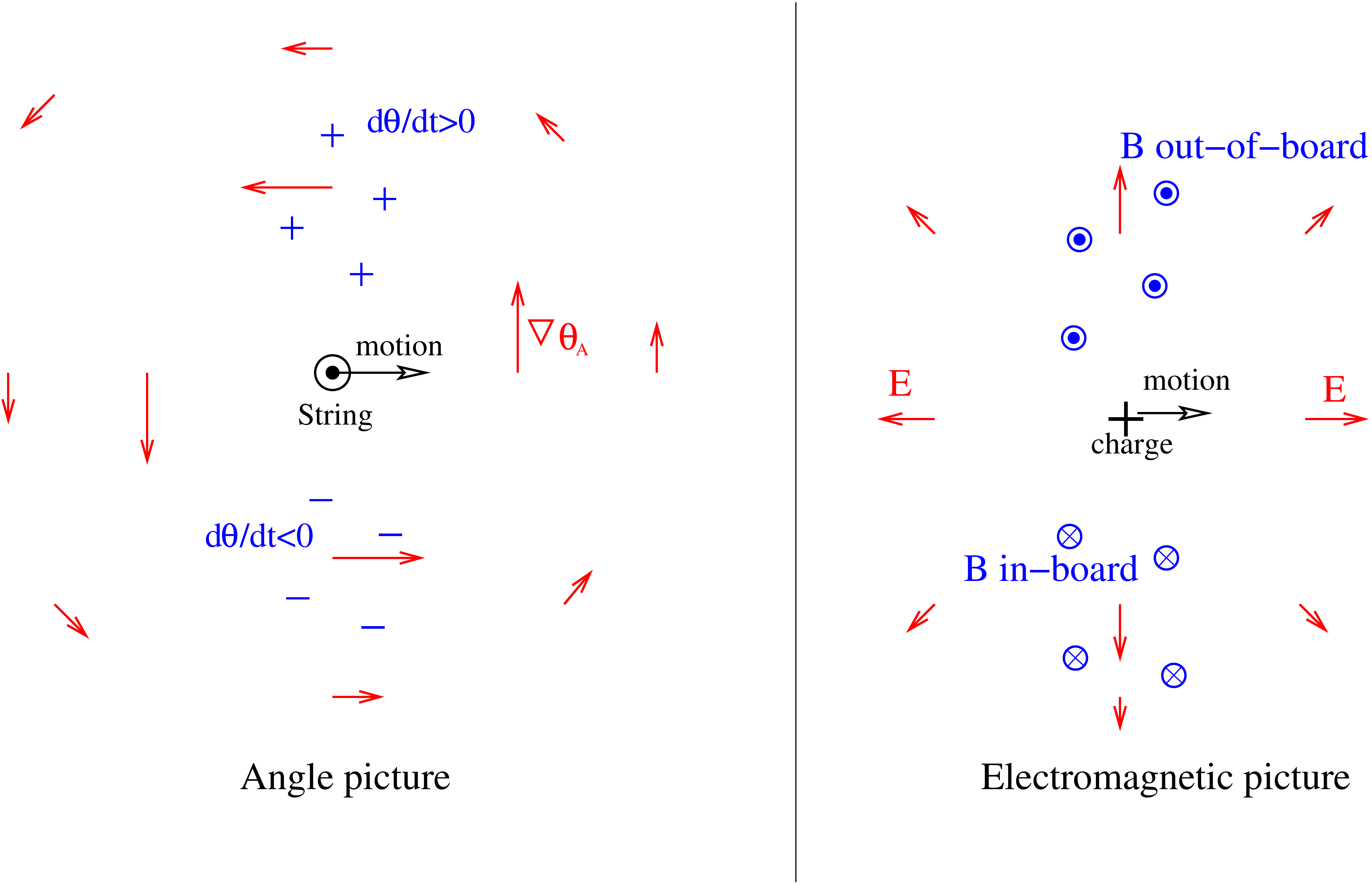}}
  \caption{\label{fig:dualpic}
    The dual pictures:  a moving string in terms of angles (left) and
    in terms of electromagnetic fields (right).}
\end{figure}

The electric flux through a loop enclosing a positive-vorticity string
is
\be
\label{Echarge}
\oint E_i \hat{n}_i dx \equiv \oint \epsilon_{ij} E_i dx_j
 = \oint f_a \partial_i \tA dx^i = 2\pi f_a
\ee
so the strings act as electric charges of charge $\pm 2\pi f_a$.
Therefore axion field dynamics in 2+1D are equivalent to
electrodynamics of particles with charge $\pm 2\pi f_a$.  If we include
all scales down to the string core scale, the electric charges' mass
arises entirely from the field (self-)energy.  But if we regulate
the charges' self-energies at a scale $r_0$, then we can include the
effect of very small string cores by giving the charges masses,
$M = \pi f^2_a \ln(m_s r_0)$, with $m_s$ (again) the mass of the radial
excitation or the inverse of the string core size.
The electromagnetic duality for a string of positive winding number is
illustrated in Figure \ref{fig:dualpic}.

\subsection{Regulation by smeared charges}

The duality to 2+1 dimensional electromagnetism suggests that string
cores can be treated as charged particles in
a ``particle-in-cell'' approach \cite{PIC}.  To achieve a
core size of $a e^{-\kappa}$, with $a$ the lattice spacing and $\kappa\sim 70$,
we could choose to make the charges have mass $M=\pi f^2_a \kappa$.
However, pointlike charges turn out to be problematic when interacting
with a lattice electromagnetic field.  The highest wave-number $k$
modes of the lattice show Lorentz-violating unphysical behavior, and
we must protect the point-charges from interacting with them through a
form-factor, which is achieved by smearing the charge into a ball.
Alternately, in the continuum we can view this as an explicit
mechanism to regulate the UV contribution to the string's self-energy.
The network dynamics will only be well
represented on scales larger than the radius of this ball.  We will do
our best to add in any physics which is disturbed by this ball, such
as the interactions between strings when they get very close together.
But since the final axion number is dominated by long-wavelength
excitations, the regulation should not significantly impact our
results for axion number generation.

We will show how to smear the charges in the continuum, and then find
an implementation of the resulting equations on the lattice.
The implementation we use will not be fully covariant; the ball's
shape will not respond to Lorentz contraction and will respond
instantly to changes in the charge's velocity, without retardation.
These are minor problems in 2+1D if the charges are heavy, because
they will then be nonrelativistic and their accelerations will be
small.

To begin the implementation, we pick the charge distribution within
the ball.  Specifically, for a positive-charge string at position $y$
we choose the charge density at $x$ to be $\rho(x) = f_a g(|x-y|)$ with
\be
\label{gx}
\int g(x) \; d^2 x = 2\pi \int r g(r)\; dr = 2\pi \,,
\qquad
g(r>r_0) = 0
\ee
so the charge has compact support.  We will also choose $g(r)$ to go
continuously to zero as $r\to r_0$, so the charge density is
continuous.%
\footnote{%
  In our numerical implementation we use
  $g(r) = 4(r_0^2-r^2) r_0^{-4} \Theta(r_0-r)$, so
  $f(r<r_0) = (1-r^2/r_0^2)^2$.}  
We will also introduce
\be
\label{fr}
f(r) \equiv \int_r^\infty r' g(r') \; dr'
\ee
which is the charge fraction lying outside radius $r$.  It obeys
$f(r>r_0) = 0$, $f(0)=1$, and $df/dr = -r g(r)$.

For ease of presentation we will consider a single positive-charge
string.  The case of many strings just involves a summation and $\pm$
signs.  The charge density modifies the Maxwell equations;
\be
\label{max2}
\partial_\mu F^{\nu\mu}(x) = J^\nu = f_a g(r) v^\nu
\ee
where $r_i = x_i-y_i$ is the vector from the charge's location $y_i$
to the position of interest $x_i$ and
$v^\nu = (1,dy_i/dt) = (1,-dr_i/dt)$.

We need to find a modified set of $\tA$ field dynamics and a modified
relation between $F_{\mu\nu}$ and $\tA$ under which \Eq{max2} is
satisfied.  There are two reasons to prefer to work in terms of
$\tA$.  First, it is a more compact way of writing the physical
degrees of freedom.  Second, the symmetry-breaking potential is only
known in terms of $\tA$.  We will modify \Eq{defF} by expressing it in
terms of a ``covariant'' derivative of $\tA$:
\be
\label{Adef}
F^{\mu\nu} = -f_a \epsilon^{\mu\nu\alpha} D_\alpha \tA \,, \qquad
D_\alpha \tA = \partial_\alpha \tA - A_\alpha \,.
\ee
We will discuss the physical interpretation of $A_\alpha$ in a
moment.  To ensure that Faraday's law still holds, we need to change
the $\tA$ equation of motion:
\be
\label{newEOM}
0 = \epsilon_{\mu\nu\alpha} \partial^\alpha F^{\mu\nu}
= 2 f_a \partial^\alpha D_\alpha \tA \,,
\ee
which are the equations of motion arising from the Lagrangian
\be
\label{newL}
\LL_{\tA} = -\frac{f_a^2}{2} D_\mu \tA D^\mu \tA \,.
\ee

Now we need to determine what choice for $A_\mu$ will ensure \Eq{max2}
(Coulomb's and Ampere's laws) are satisfied.  We have
\be
\label{newMax}
\partial_\mu F^{\nu\mu} = 
- f_a \partial_\mu \epsilon^{\nu\mu\alpha} D_\alpha \tA = J^\nu \quad
\Rightarrow \quad
 \epsilon^{\nu\mu\alpha} \partial_\mu A_\alpha = g(r) v^\nu \,.
\ee
The correct choice for $A_\mu$ to make this work is
\be
\label{Aguess}
A_i = -\epsilon_{ij} \frac{r_j}{r^2} f(r) \,, \quad
A_0 = \epsilon_{ij} \frac{v_i r_j}{r^2} f(r) \,, \quad \mbox{or} \quad
A_\mu = \epsilon_{\mu\nu\alpha} \frac{v^\nu r^\alpha}{r^2} f(r) \,.
\ee
Note that $\epsilon_{\mu\nu\alpha} v^\nu r^\alpha/r^2$ is the
derivative of the $\tA$ field around an isolated string $\tAstr$.
In other words, our choice for $A_\mu(x)$ is
$A_\mu(x) = f(r) \partial_\mu \tAstr(x)$.  Therefore
\be
\label{whyitworks}
\epsilon^{\nu\mu\alpha} \partial_\mu A_\alpha
 = \epsilon^{\nu\mu\alpha} \left( f(r) \partial_\mu \partial_\alpha \tAstr
 + [\partial_\mu f(r)] \partial_\alpha \tAstr \right) \,.
\ee
The first term vanishes by antisymmetry/symmetry on the indices
$\mu,\alpha$, while a little work and the relation
$\partial_r f(r) = -r g(r)$ confirms that the second term
reproduces \Eq{newMax}.

So the electromagnetic theory with smeared charges with smearing
charge density $g(r)$ is equivalent to the theory of angles with
$\partial_\mu \tA$ replaced by $D_\mu \tA = \partial_\mu \tA - A_\mu$
and $A_\mu$ given in \Eq{Aguess}.  The relation between the
charge density $g(r)$ and the modifier function $f(r)$ is
$f(r) = \int_r y g(y) \; dy$ or $f'(r) = -r g(r)$.

Now we turn to the string's motion.  The total Lagrangian for the
system should be $L = L_{\mathrm{str}} + \int d^3 x \: \LL_{\tA}$,
with $L_{\mathrm{str}} = -M\sqrt{1-v^2}$ the standard Lagrangian for a
relativistic massive point particle.  Varying with respect to the
string's position, one finds
\be
\label{stringEOM}
\partial_t \left( \frac{Mv_i}{\sqrt{1-v^2}} \right) = F_i \,,
\ee
where the force arises from the dependence of $\LL_{\tA}$ on the
string's location.  We find the expected Lorentz force law,
\be
\label{Lorentz1}
F_\mu = \int d^2 x\: \rho(x)\, F_{\mu\nu}(x) v^\nu
\ee
with space component
\be
\label{Lorentz2}
F_i = \int d^2 x \: \rho(x)\, \left( E_i + \epsilon_{ij} v_j B \right)
= f_a^2 \int d^2 x \: g(|x-y|) \left( \epsilon_{ij} D_j \tA
+ \epsilon_{ij} v_j D_t \tA \right) \,.
\ee
The time component $F_0$ determines how fast the string and the field
exchange energy.  The form of \Eq{Lorentz1} ensures that
$F^0 = v_i F_i$ as expected.

Let us pause to interpret these equations and in particular the role
of $A^\mu$. Far from a string, $A^\mu=0$ and the equations of motion
are as usual, $\partial_\mu \partial^\mu \tA = 0$.  But near the
center of the charge ball, where $f(r) \simeq 1$, the $A_\mu$ term
cancels $\partial_\mu \tA$ provided that $\tA$ takes the form of the
field near a string core, $\tA=\tAstr$.  So the energy is minimized by
having $\tA\simeq \tAstr$ in the interior.  In particular, this forces
a singularity onto $\tA$ at $r=0$; the energy is only finite when
$\tA$ possesses this singularity.
In other words, when $\tA$ takes the cosmic-string form,
the gradient energies associated with its spacetime variation at
distances $r < r_0$ are cut off, and associated instead with the
explicit string mass.

We need to choose $M$ such that it incorporates the energy in the
string from all scales $r < r_0$, where the $|D_i \tA|^2$ terms have
been cut off.  According to \Eq{kappa}, the string's energy (per
length in 3+1D) is $f_a^2 \pi \int_{1/m_s}^{H^{-1}} dr/r$.  The $\tA$
field gradients will capture the $f_a^2 \pi \int_{r_0}^{H^{-1}} dr/r$
part of this energy, so the mass $M$ is required to capture the rest:
$M = \pi f_a^2 \int_{1/m_s}^{r_0} dr/r = \pi f_a^2 \ln(r_0 m_s)$.
We can think of this as a matching condition that the charge-ball
treatment correctly captures the energy and inertia of the string.

When there are more than one string, each string responds to the $\tA$
field gradients caused by the other strings' presence, which induces
inter-string interactions.  Because the analogy to the electromagnetic
theory is exact, these interactions are the same as the
electromagnetic forces between charge balls.  Since the theory is
relativistic and has radiation fields, orbitally bound pairs of
(oppositely-charged) strings tend to inspiral and annihilate.
However, under the charge-ball description we have built, this
inspiral will not proceed correctly. While strings at separation
$R > 2r_0$ feel the usual Coulomb force between point charges,
$\vec{F}_{\mathrm{pt.chg}} = \pm 2\pi f_a^2 \vec{R}/R^2$ (in the
nonrelativistic limit), when the charges get close together, the balls
overlap and the strength of the interaction is reduced.  It is
important to re-introduce the missing Coulombic interaction, lost due
to the overlap of the charge balls.  Otherwise, the charges' inspiral
and annihilation will not proceed.  Define $h(R)$ as the fraction
by which the Coulomb interaction of overlapping balls is smaller than
that between point charges:
\be
\label{defh}
h(R) = \frac{|F_{\mathrm{pt.chg}}| - |F_{\mathrm{ball.chg}}|}
{|F_{\mathrm{pt.chg}}|} \,.
\ee
Then we should add an explicit force between nearby charges, of
magnitude $h(R) F_{\mathrm{pt.chg}}$.  This procedure is not exact; it
is only correct in the nonrelativistic limit.  But this is actually a
good approximation for large $M/f_a^2$, and in any case we only need to
include the short-distance interactions roughly to ensure that the
inspiral and annihilation proceeds.

When the
strings pass nearer still, the replacement of point charges with balls
also reduces their tendency to radiate, which requires an added
radiation-reaction force.  We discuss this radiation-reaction force in
more detail in Appendix \ref{apprad}.  With both forces included, we
at least semi-quantitatively reproduce the physics of inspiral and
annihilation.

Note however that if radiation is included by a radiation-reaction
force rather than by explicit interactions with the $\tA$ field, the
radiated energy and any associated axion number is lost to the
system.  This is a problem if our goal is to track the total energy in
the $\tA$ field, but not if our goal is to determine the number of
axions.  The strings' mass-energy $2M$ is liberated by the inspiral
process at small separation, with equal energy released in each equal
logarithmic range of wave number.  However, the axion number
associated with an excitation of energy $E$ and wave-number $k$ is
$\nax = E/\omega_k \simeq E/k$, which becomes insignificant at large
$k$.  Therefore, from the point of view of tracking axion number in
the 2+1D theory, it is not important to account for radiation after a
binary pair of strings becomes tightly bound.

\section{Lattice implementation}
\label{latt}

\begin{figure}[h]
  \hfill
  \epsfxsize=0.75\textwidth\epsfbox{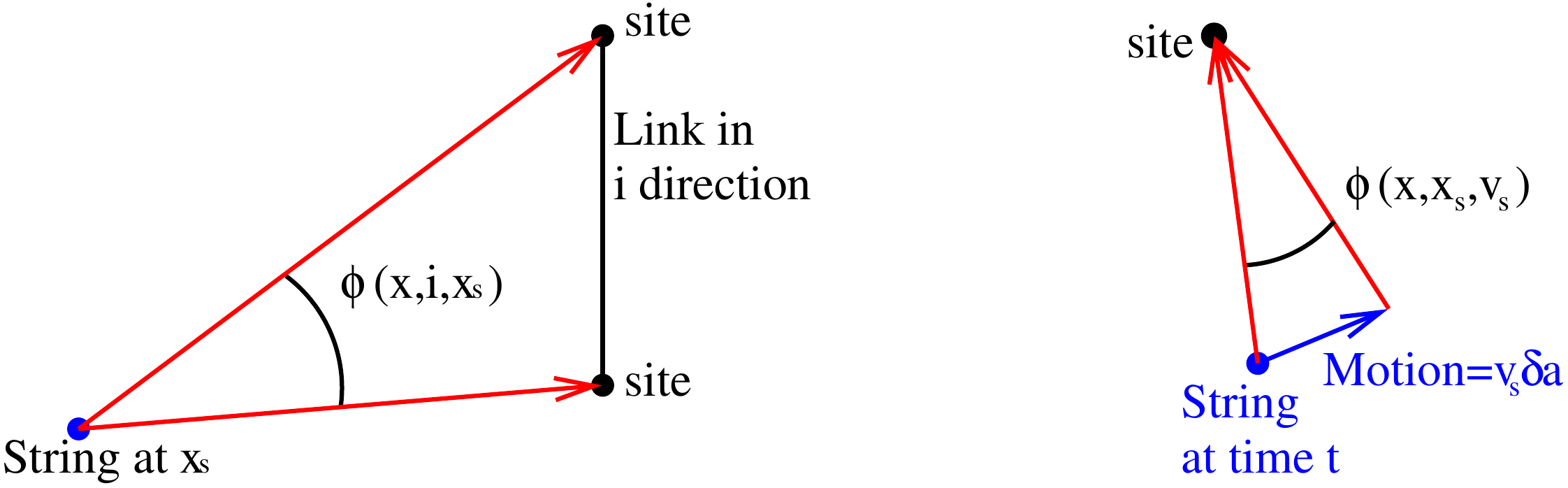}
  \hfill $\vphantom{.}$
  \caption{The angles $\phi(x,i,x_s)$, left, and $\phi(x,x_s,v_s)$,
    right.
    \label{figphi}}
\end{figure}

Now we present an implementation of these particle-in-cell equations
on the lattice.  The possibility of such an implementation is a key
result of this paper.  But readers who are not interested in such
details can skip this section.  We define the field $\tA$ on the sites
of a 2D square lattice with spacing $a$, while the strings are taken
to have continuous positions.  Time must also be discretized, with
spacing $\delta a$, $\delta \ll 1$.  On each lattice link we define
$D_i \tA$ as
\be
\label{Dilatt}
D_i \tA(x) = \tA(x+a\hat{i}) - \tA(x) - A_i(x) \,,
\ee
and on the temporal link we define
\be
\label{D0latt}
D_0 \tA(x,t) = \tA(x,t+\delta a) - \tA(x,t) - A_0(x,t) \,.
\ee
The standard meaning for these derivatives would include additional
factors of $1/a$ and $1/\delta a$.  Note that $D_i \tA(x)$ really
``lives'' at $x+a\hat{i}/2$ halfway along the link, while
$D_0 \tA(x,t)$ really ``lives'' at time $(t+\delta a/2)$.
Also, the $D_\mu \tA$ are only defined modulo $2\pi$; we always take them%
\footnote{%
  Numerically, we implement both as integers; $\tA \in [0,2^{29})$ are
  unsigned and $D_\mu \tA \in [-2^{28},2^{28})$ are signed, with bit
  masks used to enforce periodicity.}
in the interval $[-\pi,\pi]$.  The field update rule is
\bea
\label{lattfieldEOM}
D_0 \tA(x,t) &=& \frac{(t-\delta a/2)^2}{(t+\delta a/2)^2}
D_0 \tA(x,t-\delta a)
\\ && {} \nn
+ \frac{t^2 \delta^2}{(t+\delta a/2)^2}
\left[ \frac{a^2 \chi(T)}{f_a^2} \sin \tA + \sum_{i=1,2}
 \left( D_i \tA(x,t) - D_i \tA(x-a\hat{i},t) \right) \right]\,.
\eea
The factors of $t^2$, $(t\pm \delta a/2)^2$ incorporate Hubble
expansion; the power 2 is because we are in conformal time in a
radiation dominated universe.%
\footnote{%
  Using conformal coordinates has two other effects.  First,
  $\chi(T)/f_a^2 = m_a^2$ must be rescaled into conformal coordinates.
  If $\chi(T) \propto T^{-n}$ in physical units, then in conformal
  coordinates $a^2 \chi(T)/f_a^2 \propto T^{-n-2} \propto t^{n+2}$.
  We return to this point at the start of Section \ref{results}.
  Second, the string mass $M=\pi \kappa = \pi \ln(f_a r_0)$.
  Now $f_a$ is fixed in physical units, but we keep $r_0$ fixed in
  comoving lattice units.  Therefore $M$ should slowly increase with
  time, $t \partial_t M = \pi$.  The change over the physically
  interesting part of the evolution (roughly, $t=t_*$ to $t=3t_*$) is
  small, and we have neglected this effect in this explorative study.}
We store $\tA(x,t)$ and $D_0 \tA$ and
perform this update by evaluating $D_i \tA$ on each link.  The lattice
definition of $A_i(x)$ is
\bea
\label{lattAi}
A_i(x) &=& \sum_{\mathrm{strings}} \pm f(|x+a\hat{i}/2-x_s|)
\phi(x,i,x_s) \,, \nn \\
\phi(x,i,x_s) & = & \atan \frac{(x+a\hat{i}-x_s)_y}{(x+a\hat{i}-x_s)_x}
- \atan \frac{(x-x_s)_y}{(x-x_s)_x} \,.
\eea
That is, $\phi(x,i,x_s)$ is the angle subtended by the link from
$x$ to $x+a\hat{i}$ as seen by the string at point $x_s$, as
illustrated in Fig.~\ref{figphi}.

The angle $\phi(x,i,x_s)$ is the
geometrical interpretation of
$\int_x^{x+a\hat{i}} (-\epsilon_{ij} r_j/r^2)dx$ along the length of the
link.  Note that we evaluate $f(r)$ using the separation between the
string's position $x_s$ and the center of the link.  In practice we
find $A_\mu$ by performing a single loop over all strings with a
nested loop over sites or links within a
$2r_0\times 2r_0$ box around each string.  Therefore the algorithm
scales linearly with system volume -- though the numerical cost of
finding $A_\mu$ scales as $(r_0/a)^2$ and dominates the numerical
costs at early times when there are still many strings.

The value of $A_0(x,t)$ can be evaluated after we have determined how
the string will move between $t$ and $t+\delta a$.  Defining the
string velocity between times $t$ and $(t+\delta a)$ as $v_s(t)$, the
value of $A_0(x,t)$ is
\bea
\label{lattA0}
A_0(x,t) & = & \sum_{\mathrm{strings}} \pm f(x-x_s-\delta a v_s/2)
\phi(x,x_s,v_s) \,, \nn \\
\phi(x,x_s,v_s) & = & \atan \frac{(x-x_s-v_s \delta a)_y}
    {(x-x_s-v_s \delta a)_x} - \atan \frac{(x-x_s)_y}{(x-x_s)_x}
\eea
where $\phi$ is the angle change of the charge as it moves from $x_s$
to $x_s + v_s \delta a$, as seen from the lattice site, also
illustrated in Fig.~\ref{figphi}.  Again, this
is the geometrical interpretation of
$\int_{t}^{t+\delta a} (\epsilon_{ij} v_i r_j/r^2)dt$.
Here $x_s$ and $v_s$ are both evaluated at $t$, so $g(r)$ uses the
separation at time $t+\delta a/2$.

We will scale out the overall $f_a^2$ factor and write $M=\pi \kappa$
a pure number.  The string's trajectory is determined by
\bea
\label{xsupdate}
x_s(t+\delta a) & = & x_s(t) + \delta a v_s(t) \,,
 \\
\label{vupdate}
 \frac{M v_s(t)}{\sqrt{1-v_s^2(t)}}
& = & \left( \frac{t-\delta a/2}{t+\delta a/2} \right)^2
\frac{M v_s(t-\delta a)}{\sqrt{1-v_s^2(t-\delta a)}}
+ \left( \frac{t}{t+\delta a/2} \right)^2 \delta a
    (F_{\sss E}(t) + F_{\sss B}(t)) \,,
 \\
\label{fE}
 \pm F_{i,\sss{E}}(t) &=& \sum_{x,j} \epsilon_{ij} g(|x+a\hat{j}/2-x_s|)
D_j \tA(x) \,,
 \\
\label{fB}
 \pm 2F_{i,\sss{B}}(t) & = & \frac{(t-\delta a/2)^2}{\delta t^2}
\sum_{x,j} \epsilon_{ij} v_j(t-\delta a)
g(|x-x_s+v_s(t-\delta t)/2|)  D_0 \tA(x,t-\delta a)
\nn \\ && {}
+ \frac{(t+\delta a/2)^2}{\delta t^2}
\sum_{x,j} \epsilon_{ij} v_j(t) g(|x-x_s-v_s(t)/2|)  D_0 \tA(x,t) \,.
\eea
The two components of the force are the electric force, from space
gradients of $\tA$, and the magnetic force, from its time
derivatives.  The one subtlety is that $v_i g(..)$ in the last line depends
on the final velocity, which we don't know until we have computed it.
The update is therefore implicit.  We handle this by making an initial
guess for $v_s(t)$ based on a linear trajectory, using it to evaluate the
final magnetic force term, and iteratively re-substituting the
determined $v_s$ to recompute the magnetic force.  The iteration
converges by a factor of $\delta/M \sim 10^{-3}$ per repetition.
$A_0(x,t)$ and therefore $\tA(x,t+\delta a)$ are only evaluated after
$v_s(t)$ has been found.

To include the explicit forces between strings with separation
$R<2r_0$, we sort strings into boxes $2r_0$ on a side and search for
nearby strings by comparing all string pairs in the same or
neighboring (direct or diagonal) boxes.  This approach keeps the
algorithm linear in system volume.  For each nearby string pair,
we apply an inter-string force of $\pm 2\pi h(R) R_i/R^2$, with
$h(R)$ defined in \Eq{defh}.  When
strings pass even closer, with separation $\sim v_s r_0$, the
radiative energy losses are not fully included, and we include a
radiative reaction force, as motivated in Appendix \ref{apprad}.
Lastly, we assume that any pair of strings which get closer than a
distance $\rmin \ll a$ will annihilate, and rather than following
their final inspiral we remove them from the simulation.  When
annihilating a pair of strings, we add a contribution to \Eq{lattA0}
where $\phi(x,x_s,v_s)$ is the angle difference, as seen from the
lattice site, between the two strings which will annihilate. This is
the same as incorporating the shifts to the $\tA$ fields which would
occur from sliding one string on top of the other before removing them
from the simulation.  This prevents an unphysical ``kick'' to the
fields when the strings annihilate and is especially important when
the annihilating strings happen to be very close to a lattice site.

Our choice for initial condition is to draw $\tA \in [0,2\pi]$
randomly at each lattice site, and then to apply a few steps of
checkerboard nearest-neighbor smearing.%
\footnote{One checkerboard step is to replace $\tA(x)$ on all
  ``odd-checkerboard'' sites with the average $\tA$ value of the 4
  nearest neighbors, defined as the unit-circle projected position of
  the centroid of the neighbor-$\tA$ positions on the unit circle.
  The next checkerboard update changes the even-checkerboard sites.}
We identify vortices in the $\tA$ field
using the algorithm presented in our previous paper \cite{axion1}, and place a
string at the center of each square with such a vortex.  Both $D_0\tA$
and $v_s$ for all strings are initialized to zero, and this initial
condition is assumed to apply at a time $t_i \geq 0$.
The specifics of the initial conditions should
not be too important since the network should converge towards a
scaling solution; we discuss this more in Appendix \ref{checks}.

It is possible for the $\tA$ field to have a vortex which is not
associated with a string, or for a string to be far from any
corresponding vortex or opposite-sign string.  In each case this would
reflect a misidentification of where strings should be based on the
$\tA$ field; either a string is missing, or an extra string was
included. That is, such errors occur when there is a failure in our
initial conditions to make the strings and $\tA$-field vortices
coincide. This occurs, at a low rate, if we use 0 or 1 checkerboard
smearings; it is exceedingly rare when we use more.  We handle it by
occasionally identifying these ``missing'' or ``orphan'' strings, and
inserting or deleting them.  All errors are caught at early times, and
this operation should be interpreted as part of the initial condition
setting algorithm.

\section{Numerical results in 2+1 dimensions}
\label{results}

It is necessary in any lattice study to ensure that the lattice
regulation is not influencing the results.  Therefore the first thing
to check is that the (unphysical) parameters of the lattice setup do
not influence results at sufficiently large scales and late
times.  The relevant parameters are $\delta$, $\rmin$, $r_0/a$, and the
specifics of our initial conditions choice.
We leave the details of these checks to Appendix \ref{checks}.
Instead we check first what the scaling solution looks like, without
the tilted potential but for different values of $M$.  Then we solve
for axion production in the case where \Eq{LLbreak} is present and
$\chi(T) \propto T^{-7}$, close to the value expected from instanton
liquid models \cite{Wantz:2009mi}.  That means that the physical
axion mass scales with conformal time as
$m_{a,\mathrm{phys}} \propto t^{7/2}$; but the conformal mass (the
oscillation rate of the field in terms of conformal time) scales
as one more power, $m_at_* = (t/t_*)^{9/2}$ (which defines the scale
$t_*$ where the mass starts to play a role).

As we will see, there is a complication.  For the case where the axion
mass turns on with time, the string pairs cease annihilating and
instead stay in tightly bound but surprisingly long-lived ``atoms.''
In the end, we will have to make analytical estimates for the
late-time axion production from these pairs, and cut them out from the
simulation, to get the final axion number produced.

\subsection{Scaling solution}
\label{secscale}

First we look at string networks without tilting the potential, that
is, keeping $m_a = 0$.  Our goal is to see that the scaling solution
exists, and to find the scaling of string density and velocity with
$M$.

To do so we perform simulations at a range of masses from $M=10$ to
$M=600$.  As initial conditions we use 2 checkerboard smearing steps
and an initial time $t_i$ between 20 and 60, with larger values for
larger $M$ so that the string density starts near its scaling limit.
We read out the string density at a final time $t=1024a$ ($t=2048a$ for
the largest $M$ value).  Other parameters are set as described in
Appendix \ref{checks}.

\begin{figure}[htb]
  \epsfxsize=0.47\textwidth\epsfbox{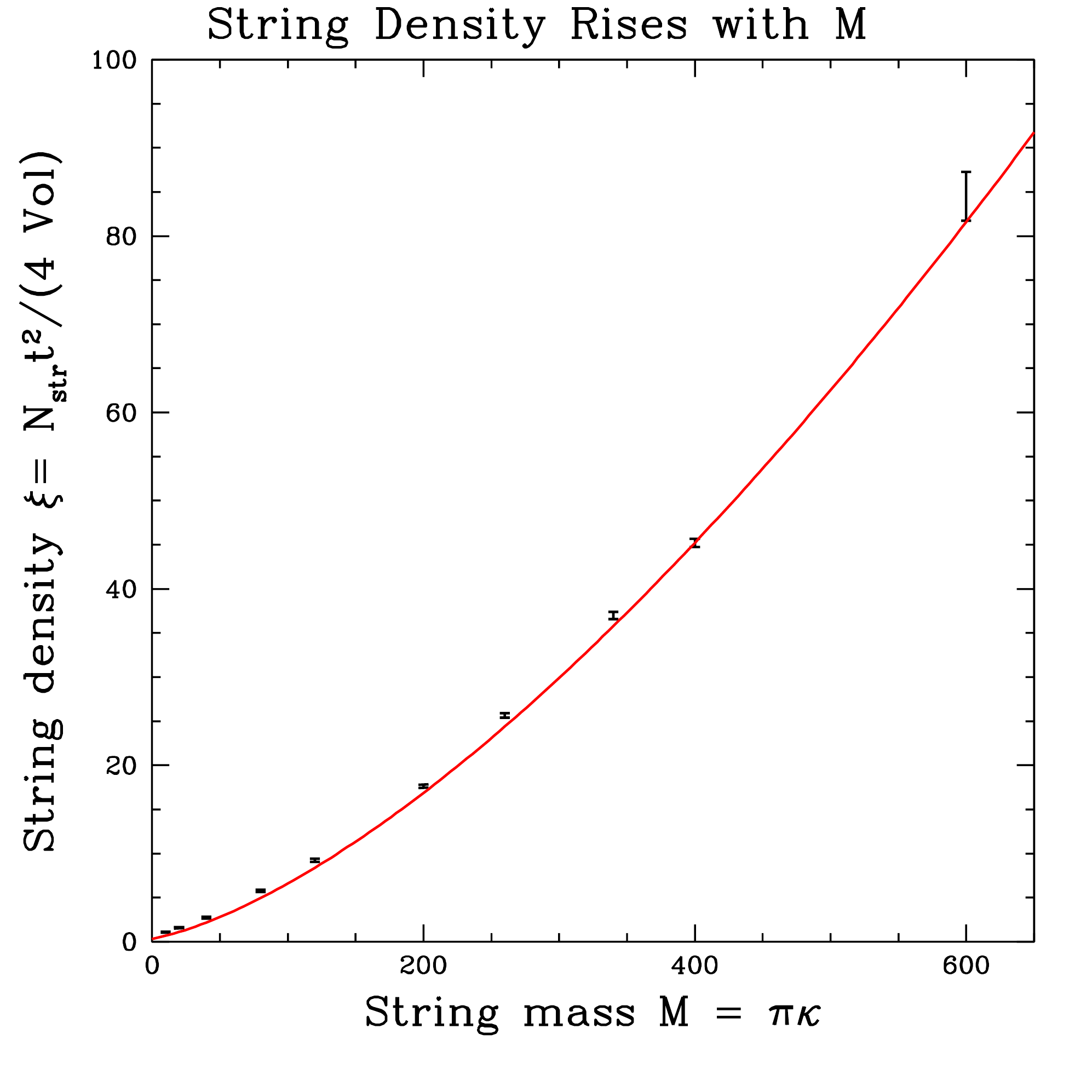} \hfill
  \epsfxsize=0.47\textwidth\epsfbox{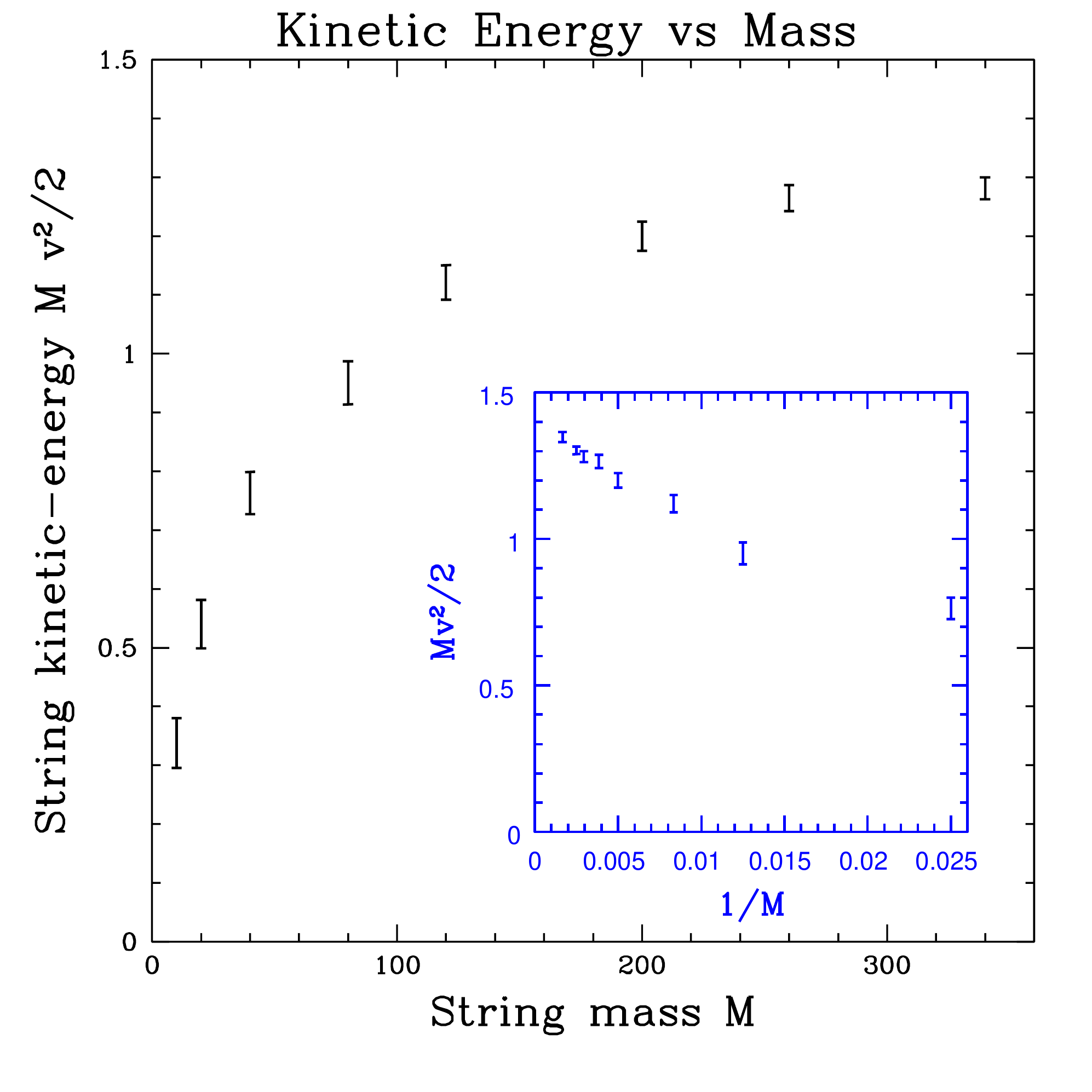}
  \caption{\label{fig:Mdepend}
    Density of strings (left) and mean kinetic energy in strings
    (right) as a function of the string mass $M$, at time $t=1024$ in
    lattice units.  The red line on the left shows
    $(M+15)^{3/2}$ behavior.}
\end{figure}

Figure \ref{fig:Mdepend} shows the dependence of the string density
and string velocity (plotted as $Mv^2/2$) on $M$. In our first paper
\cite{axion1} we make a parametric argument that $Mv^2$ should be
nearly $M$ independent, and that the string density should scale as
$M^1$.  We see that the first prediction is broadly correct.  However,
for the smallest $M$ values, much of the string's energy resides in
gradients in the $\tA$ field which are not included in the value of
$M$.  Therefore it is $(M+\pi \ln(r_{\mathrm{sep}}/r_0))v^2$, and not
$Mv^2$, which should be approximately constant.  Here
$r_{\mathrm{sep}}$ is the mean inter-string separation.  This explains
why the smallest $M$ values do not follow the expected trend.  But for
large $M$ values, where the scaling argument should be more secure,
the behavior is as expected.

On the other hand, the string density definitely does \textsl{not}
scale linearly with the string mass $M$.  Instead it rises as a larger
power, roughly $M^{3/2}$.  The red curve in the figure illustrates
$(M+15)^{3/2}$ behavior, with $15\sim \pi\ln(r_{\mathrm{sep}}/r_0)$.
This is clearly a much better fit than a straight line.
We believe that this difference is because the argument in
\cite{axion1} assumed that the most numerous strings are those which
have not become bound into tight orbital pairs.  It estimated the
density of such strings based on the time scale for two strings to get
close to each other, and then simply assumed that they will quickly
annihilate.  But we show in Appendix \ref{apprad} that this is not so;
the time it takes a bound pair of strings to inspiral and annihilate
is $t_{\mathrm{inspiral}} \sim R M^{3/2}$.  Assuming
$R\sim t M^{-1/2}$ as in the scaling solution in \cite{axion1}, the
inspiral time is $\sim tM$, long compared to the system age.  Therefore
the inspiral process takes much longer than the process for strings to
find each other, and most strings at any time are those which have
bound off in pairs and are inspiraling, not strings ``at large'' as we
assumed in \cite{axion1}.  This behavior is clear on visual inspection
of the location of strings in a simulation:  as an example, we plot
the string locations for part of the volume of an $M=400$ run at
$t=1024a$ in Figure \ref{fig:pairs}.
Tightly bound string pairs obey a
Virial relation, $Mv^2 = \pi$, as shown in the appendix.  And indeed,
in the inset of Figure \ref{fig:Mdepend} it appears that
$Mv^2/2$ asymptotes to $\sim 1.4$, somewhat below but in
reasonable agreement with this relation.

\begin{figure}[tb]
  \epsfxsize=0.44\textwidth\epsfbox{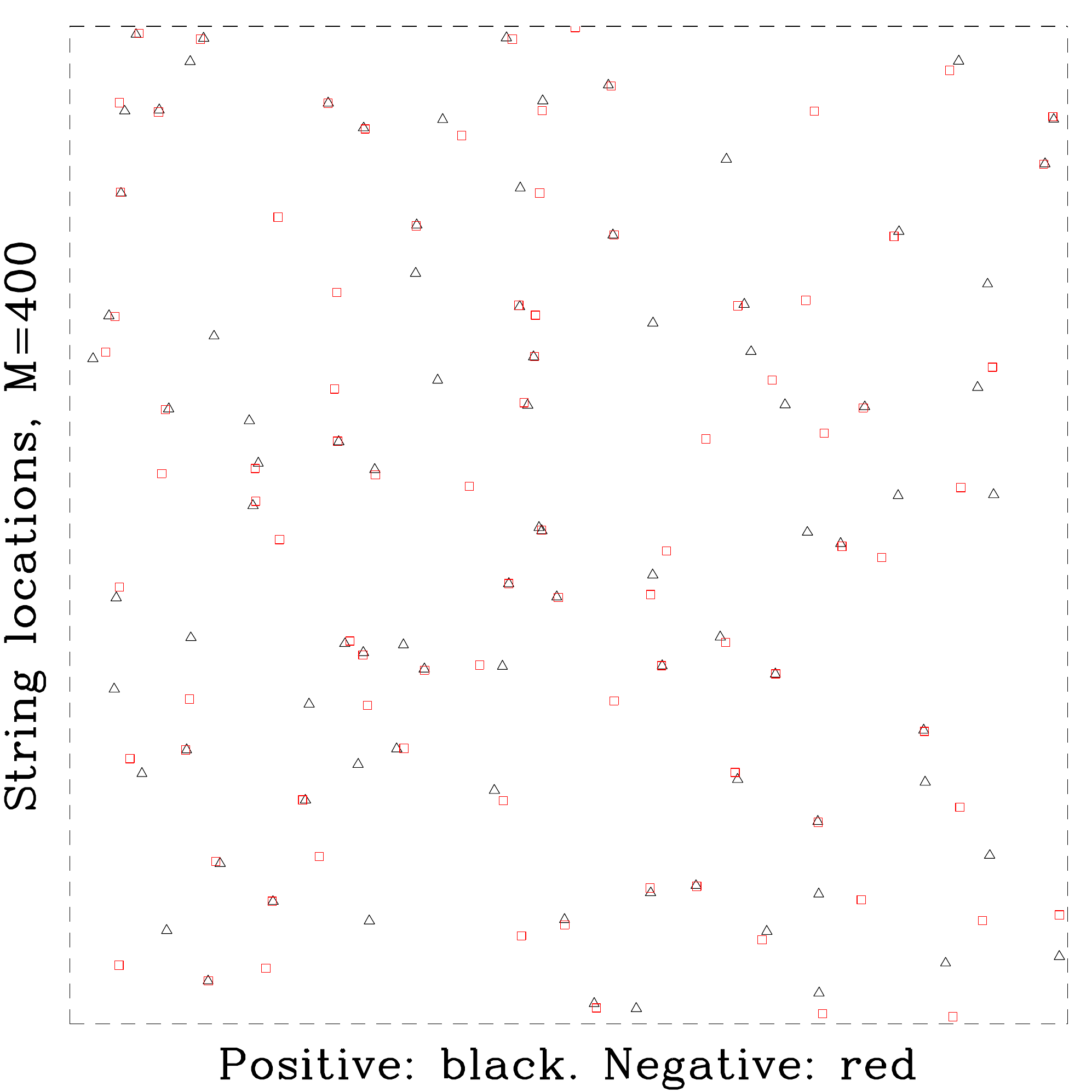}
  \hfill
  \epsfxsize=0.44\textwidth\epsfbox{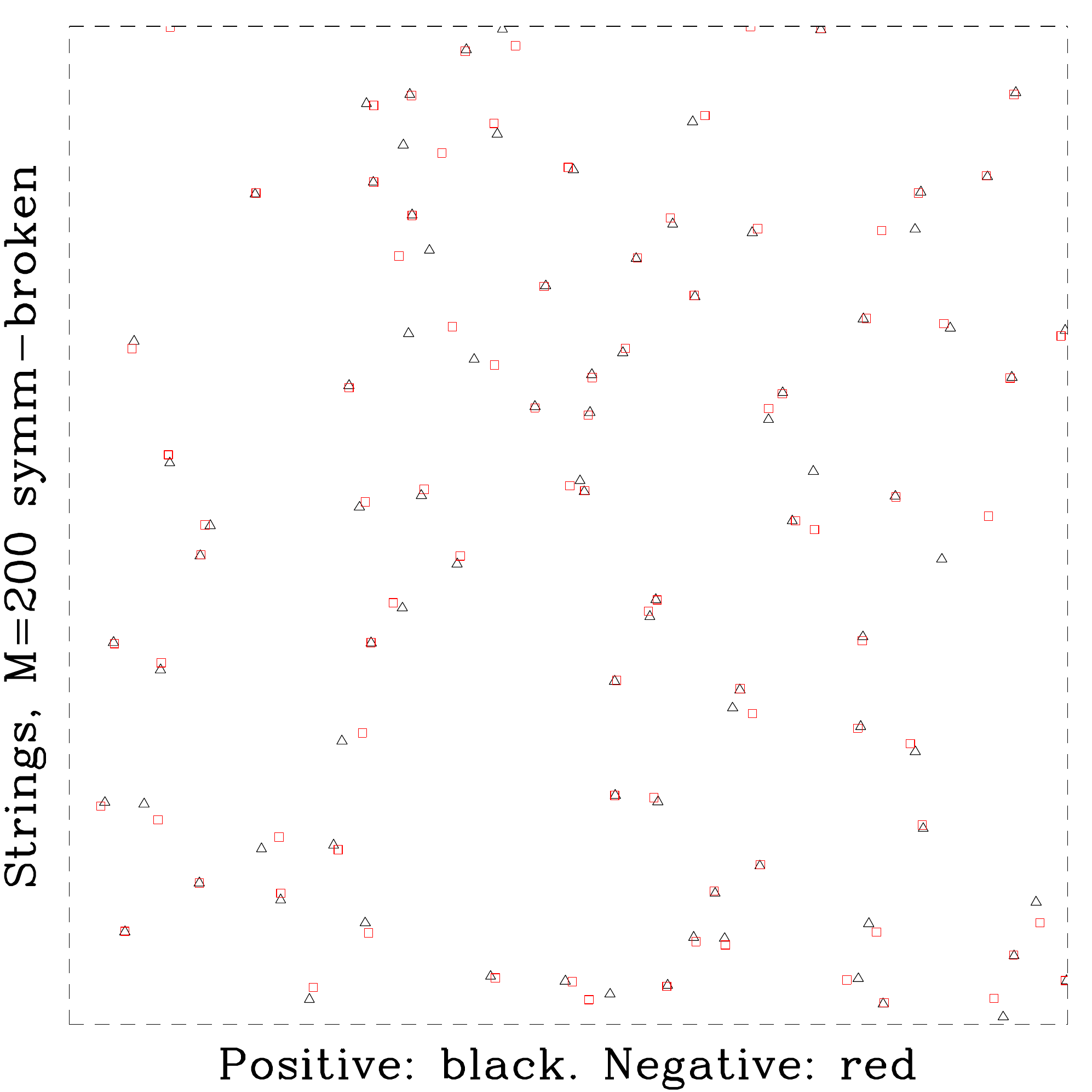}
  \caption{\label{fig:pairs}
    Location of positive (black triangles) and negative (red squares)
    strings in space:  left, for an $M=400$ simulation with no axion
    mass (untilted potential), and right, for an $M=200$ simulation
    with a tilted potential at time $t=3t_*$.  On the left, most
    strings are in associated pairs; at right, they all are, and the
    pairs are tighter.}
\end{figure}

The situation here is loosely analogous to what we expect in 3+1
dimensions.  Strings which have not bound off into pairs are analogous
to long strings, and the tightly bound pairs are like string loops.
For small $M$, radiation and inter-string attraction are important,
and string loops decay rapidly via radiation -- or pairs spiral in
quickly in 2+1 dimensions.  But for large $M$, radiation is
inefficient, and loops can persist for a long time.  Therefore the
relative importance of loops (bound pairs in 2+1D) increases with
increasing $M$.

\subsection{Axion production}
\label{secaxion}

Next we turn to axion production.  Following our previous work
\cite{axion1}, we take $\chi(T)$ to vary with temperature as
$\chi(T) \propto T^{-7}$, in which case the $\sin\tA$ term in
\Eq{lattfieldEOM} should scale with conformal time as $t^9$.  We
define $t_*$ as the time such that $m_a t_* = 1$, and we evaluate
axion number at late time and match it to the adiabatic behavior under
Hubble expansion (scaling out $f_a$ factors):
\bea
\nax &=& \half \int \frac{d^3 k}{(2\pi)^3}
\left( \sqrt{k^2+m_a^2} \langle \tA^2 \rangle
+ \frac{\langle (\partial_t \tA)^2 \rangle}{\sqrt{k^2+m_a^2}}
\right) 
\nonumber \\
& \simeq & \frac{K t_*}{t^2}
\quad \mbox{for $t\gg t_*$} \,.
\eea
As a baseline, the value of $K$ for the case where $\tA$ is uniform
and we average over the possible angle choices is $K=16$.  Our
fields-only simulations implied $K \simeq 9$, valid in 2+1D for the
small $\kappa\sim 8$ we could achieve. In 3+1D with $\kappa \sim 6$, 
fields-only simulations gave $K\simeq 8$.

\begin{figure}[htb]
  \epsfxsize=0.46\textwidth\epsfbox{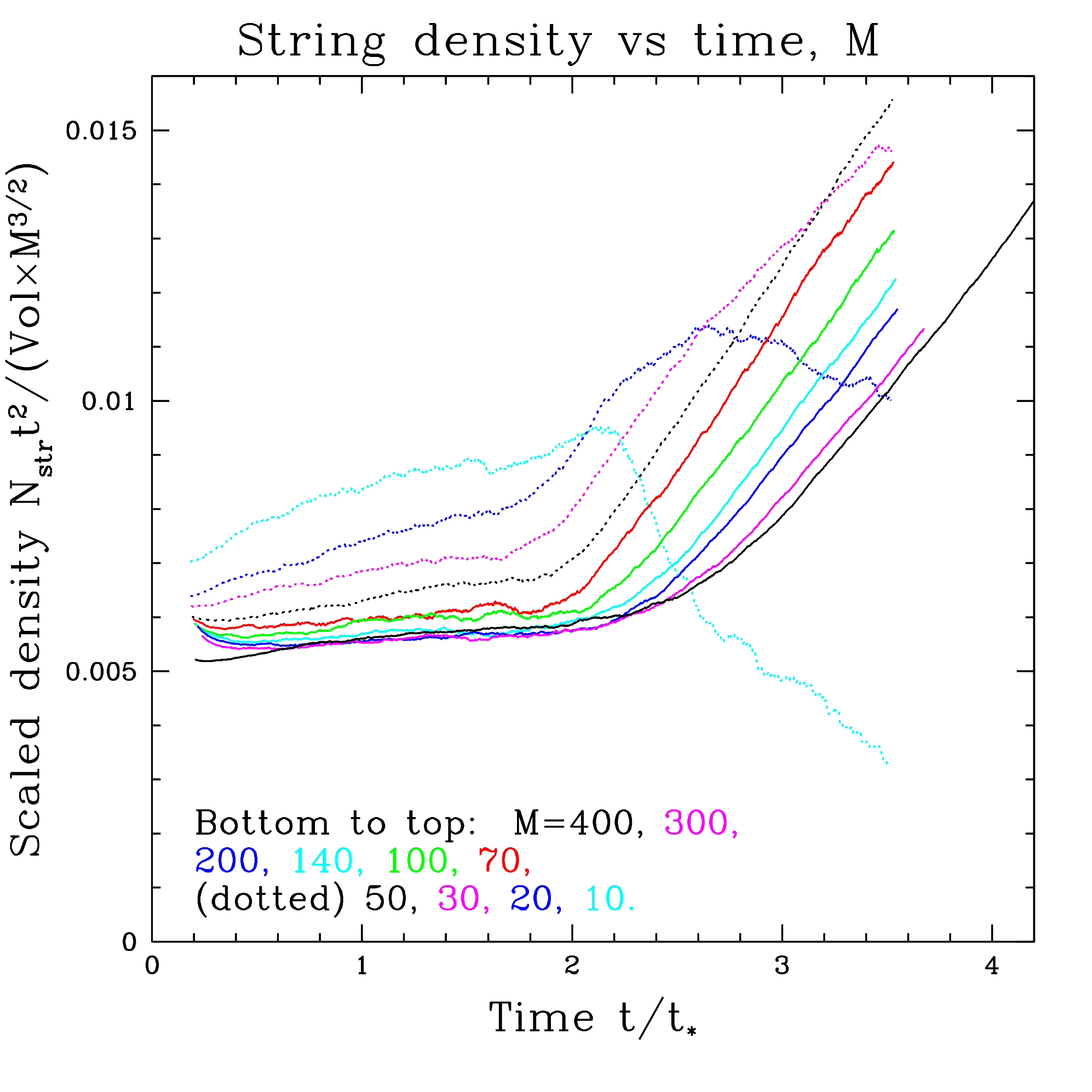}
  \hfill
  \epsfxsize=0.46\textwidth\epsfbox{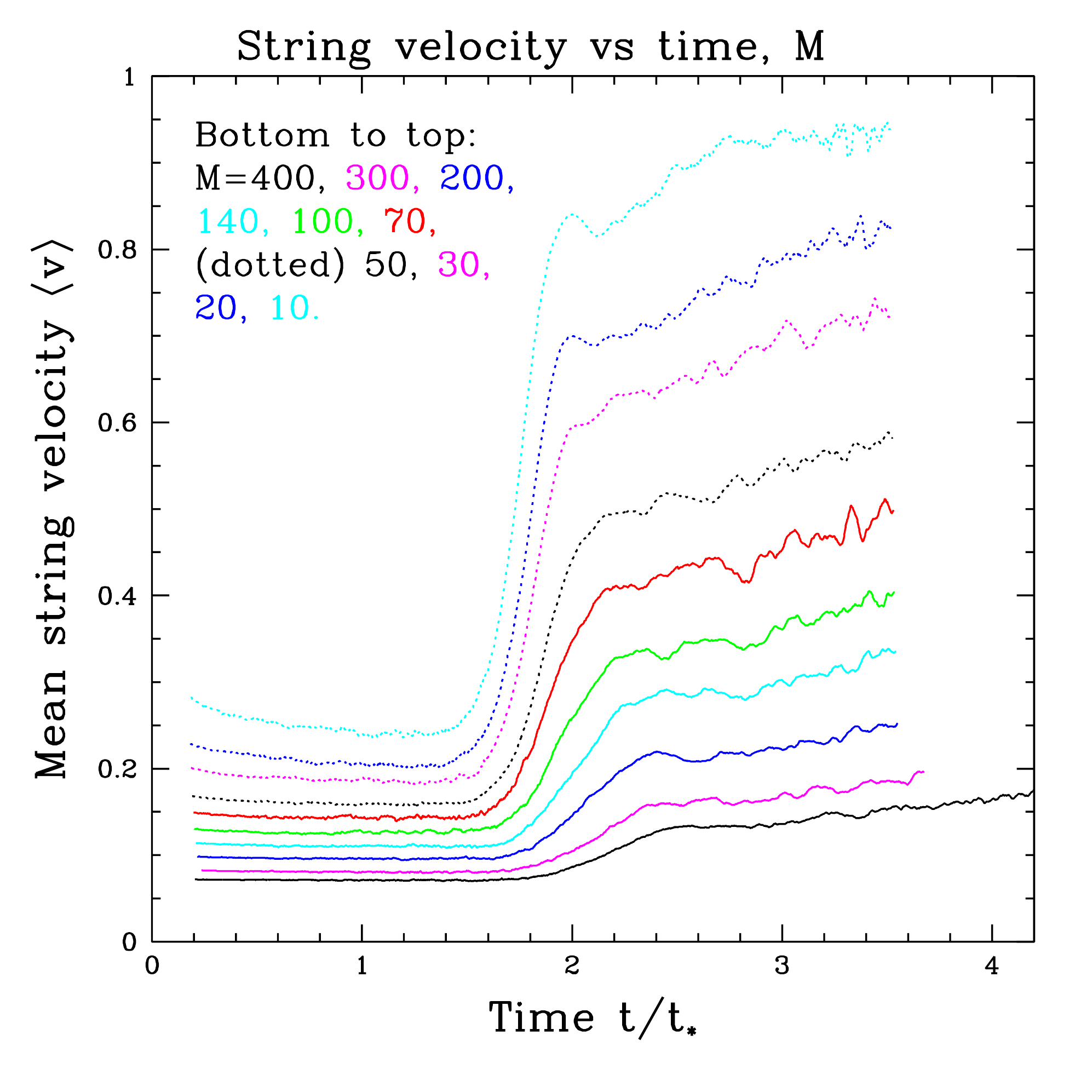}
  \caption{\label{problemfig}
    String density (left) and mean string velocity (right) as the
    effects of $m_a$ and the associated domain walls come to be felt.
    The string velocity increases, and for small $M$ the string
    density falls; but for large $M$ the string density rises relative
    to the expected $t^{-2}$ behavior which would occur in the absence
    of $m_a$ (that is, if the potential did not tilt).}
\end{figure}

It appears straightforward to determine $K$ as a function of $M$ in
our 2+1D simulations.
There is one problem, however.  The axion number should be
measured at late times when all strings have annihilated and the
fluctuations in the $\tA$ field are perturbative.  This is the
condition that $\nax$ be an adiabatic invariant, which we rely on to
relate the determined value to the value later in the history of the
Universe.  This works fine for small values of $M$.  Indeed, in our
fields-only study, we found that the strings annihilate away by around
$t=3t_*$.  But for larger values of $M$ which we can now study, such
as the value $M \simeq \pi \ln(f_a/H) \sim 200$ which we argue is
physically reasonable, the strings actually don't annihilate away, or
at least they do so over a very long time scale.  We
illustrate this in Figure \ref{problemfig}, which shows the string
density (left) and string velocity (right) as a function of $t/t_*$,
for a number of choices of $M$.  We have multiplied the string density
by a factor of $t^2$ to account for its scaling behavior, and by a
factor of $(M+15)^{-3/2}$ to scale out the $M$-dependence found in
the last subsection.  The figure shows that, at a
time around $t \simeq 1.8 t_*$ (somewhat larger for larger $M$), the
strings start to feel the added force from domain walls and pick up
speed.  This causes them to bind off into much tighter string pairs.
For small $M$ values, the strings then annihilate.  But for large $M$,
the density of strings actually falls more slowly than it would in the
absence of a potential for $\tA$.  That is, tilting the potential so
the strings attract each other more strongly actually \textsl{reduces}
their tendency to annihilate.  Therefore the curves in the figure,
which are scaled by $t^2$ to compensate for the scaling behavior,
rise.  We show that the strings are in tight pairs in the right image
in Figure \ref{fig:pairs}, which shows part of a simulation for
$M=200$ at $t=3t_*$.

The tilting of the potential causes domain walls which draw the
strings together and cause the increase in string velocity.  But it
also raises the minimum frequency of isolated axion oscillations.
This creates a mismatch between the low orbital frequency of the
string pair and the higher minimum oscillation frequency of radiated
axions, which can prevent dipole radiation.  Instead the orbital pair
can only radiate at very high, and therefore inefficient, multipole
number.  For large values $M\sim 200$ which are relevant
phenomenologically, we cannot carry out a simulation long enough for
all of the string pairs to annihilate.  Therefore we must find some
procedure to estimate the axion number which will result when the
string pairs eventually do annihilate.

Let us investigate the evolution of a bound string pair in this
regime, neglecting radiation and scattering with any other axions
present.  We will also ignore the center of mass motion of a string
pair, which is smaller than the relative motion and which redshifts
away.  The potential between the strings is well approximated by
\be
\label{stringpot}
V(r) =
8m_a r - 2\pi E_1(mr)
\ee
with $8m_a$ the domain wall tension and
$E_1(mr) = \int_r^\infty e^{-mx} dx/x$ the potential for a screened
Coulomb interaction.  We have verified
this form numerically for static strings at fixed
separation and $m_a r_0 \ll 1$.  To simplify, we will neglect the
Yukawa-like part of the potential and approximate this as
$V(r)=8m_a r$.  Under a linear potential of this form, the Virial
relation between potential and kinetic energies reads
\be
\label{Virial}
\langle V(r) \rangle =
2 \langle M v^2 \rangle \,,
\ee
where $v$ is the velocity of a string with respect to the center of
mass (half the relative velocity), so $Mv^2$ is the total kinetic
energy of the pair.   The energy evolves adiabatically as the axion
mass $m_a$ rises and as Hubble damping depletes the kinetic energy.
In our comoving coordinates,
\be
\frac{dE}{dt} = \frac{dV}{dt} + \frac{d(Mv^2)}{dt}
= \frac{dm_a}{m_a dt} \langle V \rangle
-\frac{4}{t} \langle Mv^2 \rangle
= \left( \frac{2}{3} \frac{dm_a}{m_a dt} - \frac{4}{3t} \right) E\,.
\label{stringshift}
\ee
We used that radiation-era Hubble damping gives $dv/dt=-2v/t$ in
conformal coordinates, and in the last step we used the Virial
relation.  For comparison,
the energy in long-wavelength axionic fluctuations is scaling like
\be
\label{axshift}
\frac{dE_{\mathrm{ax.fluct}}}{dt} = \left( \frac{dm_a}{m_a dt}
- \frac{2}{t} \right) E_{\mathrm{ax.fluct}} \,,
\ee
where the first term is from the adiabatic growth of the axion mass
and the second term is from Hubble drag.  Using $dm_a/dt = 9m_a/2t$,
we find that the energy in string pairs shrinks relative to
already-produced axions as
\be
\frac{E_{\mathrm{pairs}}}{E_{\mathrm{ax,fluct}}}
\propto t^{-5/6} \,,
\label{eqdilute}
\ee
so the string pairs grow less important with time.  Therefore, the
longer it takes for them to radiate away their energy into axions, the
smaller the produced axion number becomes.

We cannot follow the evolution past when $m_a a > 1$, and our
description starts to break down when $m_a r_0 > 1$.  We reach
slightly larger times by reducing $r_0$ near the end of the simulation
to keep $m_a r_0 \leq 1$, but we do not dare go beyond $r_0=2a$ for
reasons discussed in the Appendix.  So we need some technique to
estimate how much of the energy in the strings will convert into axion
number.  Our idea is to evolve the system as long as possible, and
then to remove the strings and domain walls from the simulation in a
way which leaves some of their energy behind, to capture the axion
number which they would have produced.  We have tried two approaches:
\begin{enumerate}
\item
  \label{cut1}
  When $m_a a=1/2$ we remove all strings from the simulation and
  replace values of $\tA$ close to $\pi$ with smaller values as
  follows:  for $\tA(x)\in [\pi/2,\pi]$ we apply
  $\tA \to \pi-\tA$, and for $\tA(x) \in [-\pi,-\pi/2]$ we apply
  $\tA \to -\pi - \tA$.  This keeps $\tA$ continuous and does not
  change $(\nabla \tA)^2$, though it reduces the potential energy and
  destroys all topological objects (which is the goal).  We then
  evolve the fields until $m_a a \geq 1$ so the further evolution is
  completely adiabatic, and measure axion number.
\item
  \label{cut2}
  We remove strings as above, but we remove the domain walls in a way
  which eliminates most of the energy they contain.  We leave $\tA$
  untouched in the range $[-\pi/4,\pi/4]$; in the range
  $[\pi/4,\pi/2]$ we apply $\tA \to \pi/2-\tA$ (and similar for
  negative values), and wherever $\sin(\tA)<0$ we replace $\tA$ with
  zero.  This ``cuts out'' the cores of the domain walls in a way
  which does not introduce any discontinuities in the $\tA$ field.
  This approach is almost the same as removing the strings and domain
  walls entirely, on the assumption that they will not produce any
  axions.
\end{enumerate}
\begin{figure}[h]
  \centerline{\epsfxsize=0.6\textwidth\epsfbox{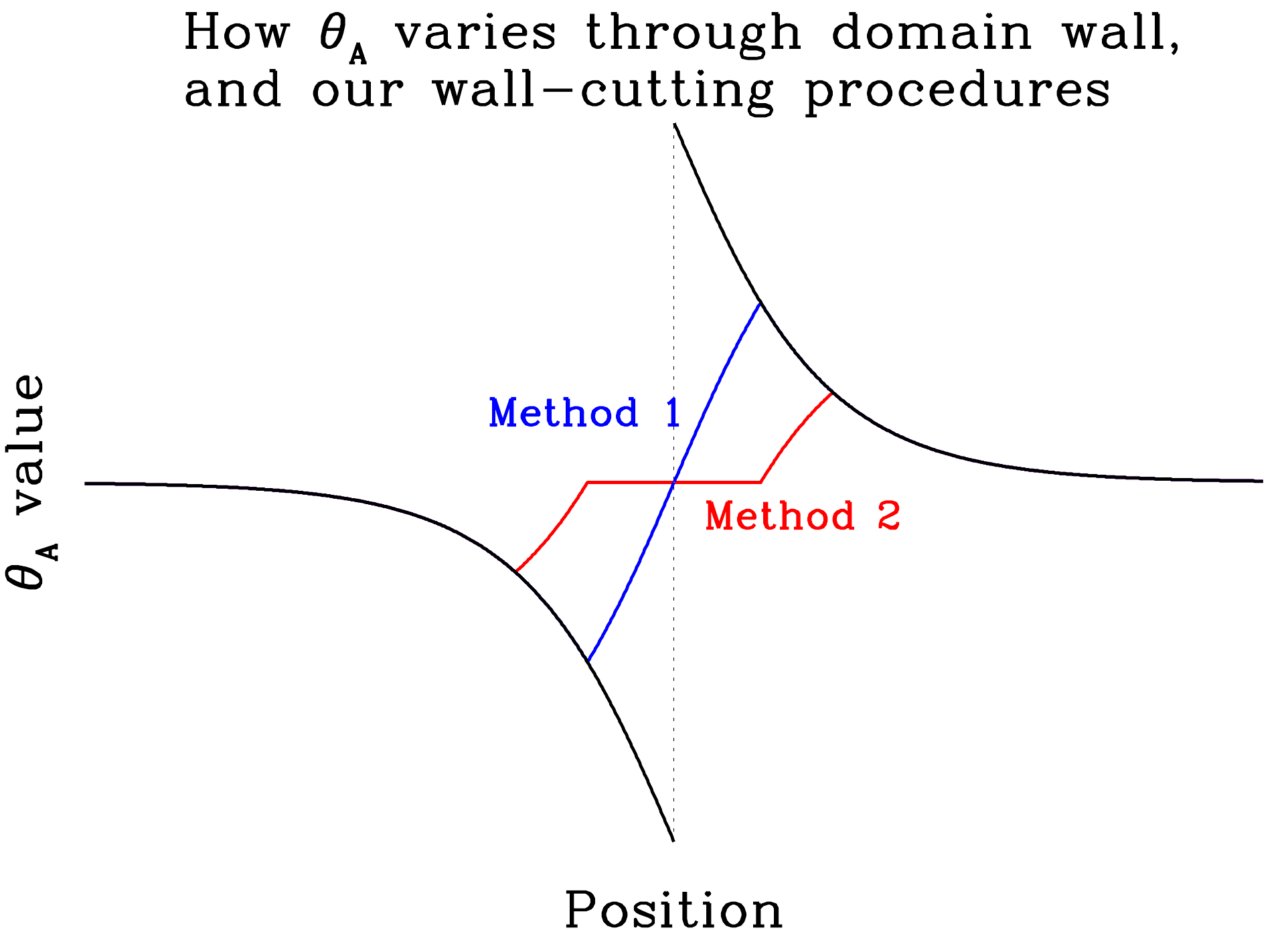}}
  \caption{\label{stringchop}
    Illustration of how $\tA$ varies through a domain wall, and how
    each of our wall-chopping procedures work.  The dotted line
    indicates periodicity, $\tA=-\pi$ and $\tA=\pi$ are equivalent.}
\end{figure}
The first approach converts most of the domain-wall energy into axions,
while the second removes most of the domain-wall energy.  They are
illustrated in Figure \ref{stringchop}.

\begin{figure}[thb]
  \epsfxsize=0.45\textwidth\epsfbox{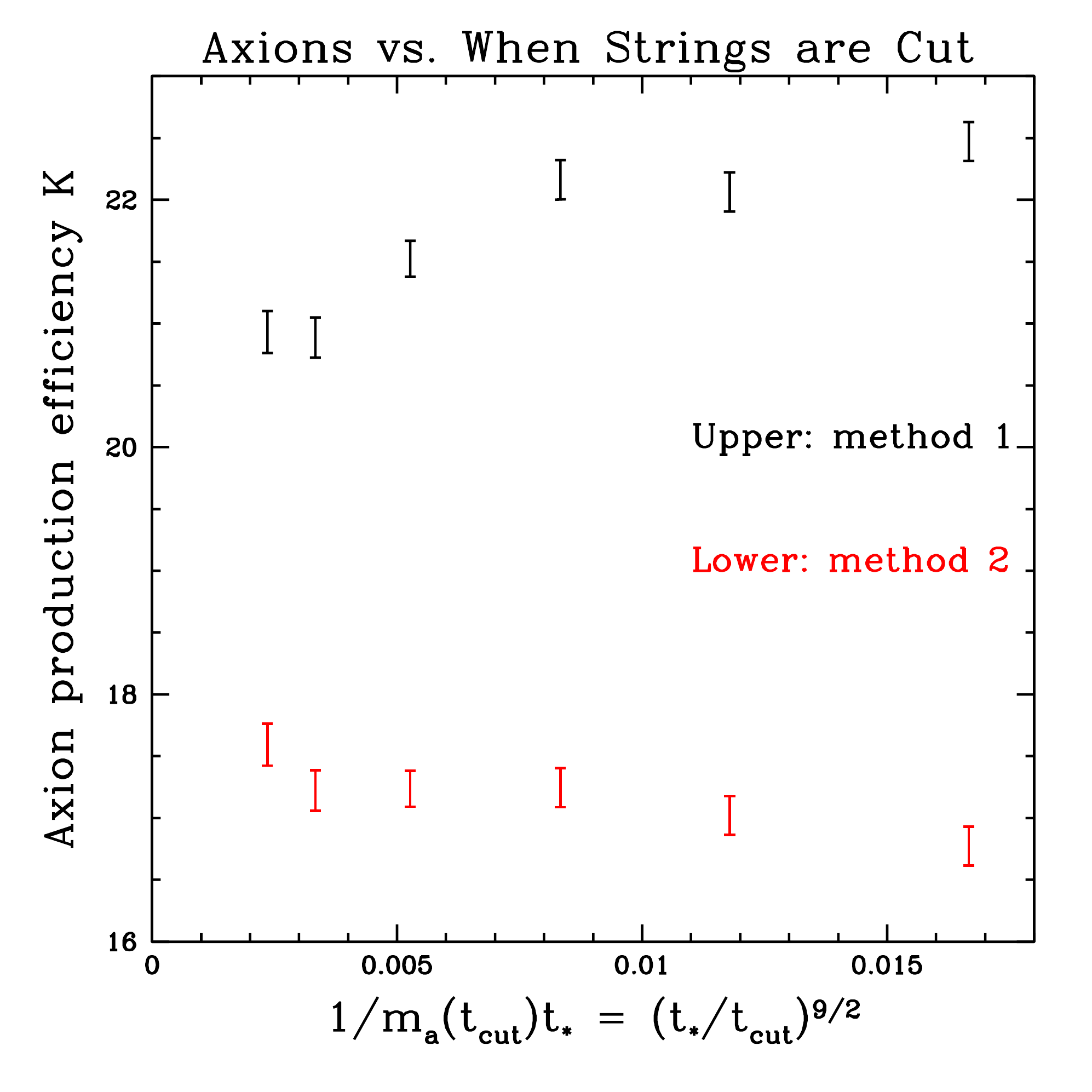}
  \hfill
  \epsfxsize=0.45\textwidth\epsfbox{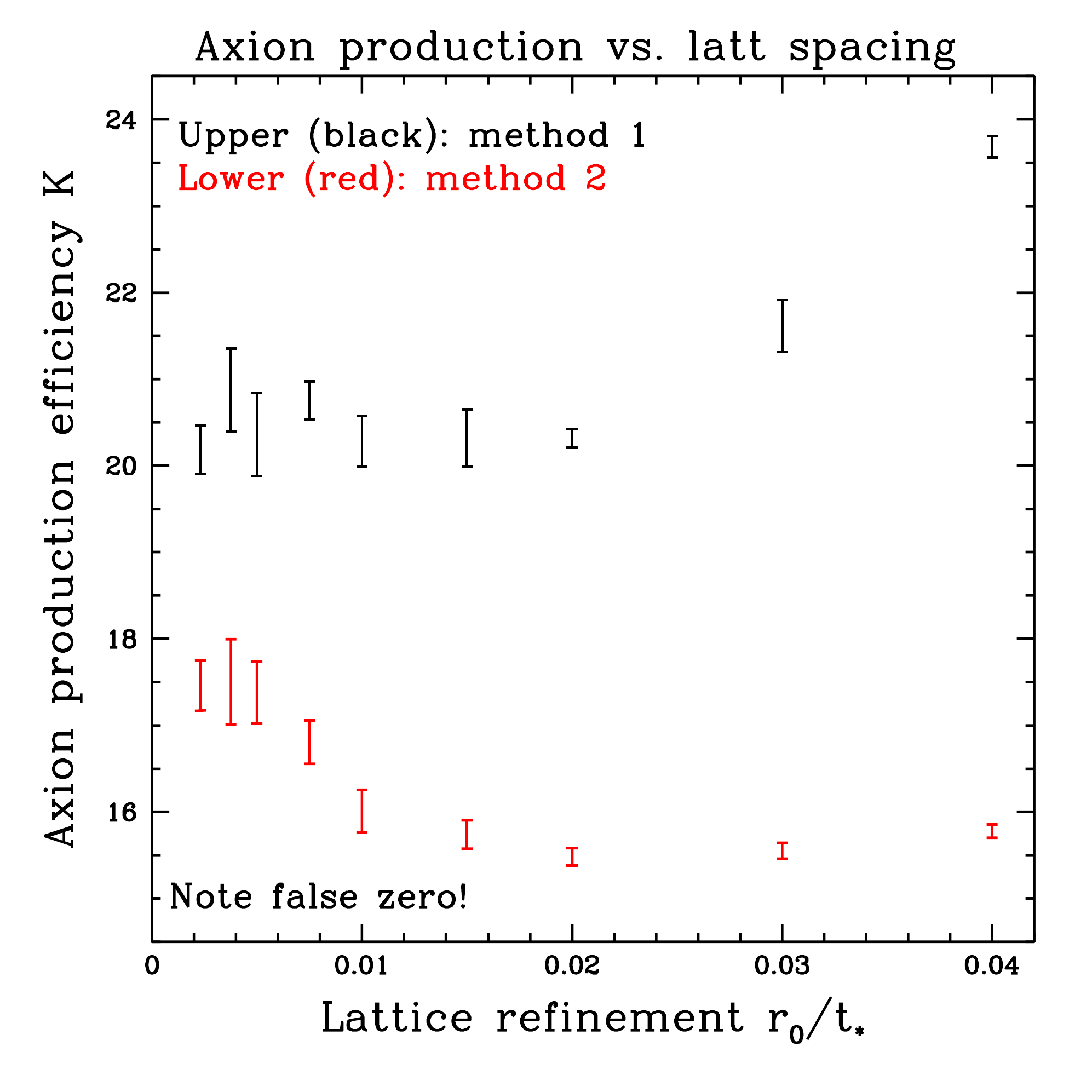}
  \caption{\label{fig:tcut}
    Axion production efficiency for $M=200$ as a function of when the
    strings/walls are cut from the simulation.  The upper (black)
    points are cutting procedure \ref{cut1}, the lower (red) points
    are procedure \ref{cut2}.  Left:  $t_*/a=600$, varying the time
    $\tcut$ when the strings are cut.  Right:  varying the lattice
    spacing $t_*/a$, always cutting at $m_a a=1/2$.}
\end{figure}

We can say something about which procedure is correct by seeing how
the produced axion number depends on the time $\tcut$ when the
string-cutting is performed.  We do this by varying the scale $\tcut$
when the cutting is performed, holding everything else fixed.  We
improve the sensitivity by correlating the statistical errors, using
the same random number seeds, so each $\tcut$ choice is applied to the
same network simulations.  The results are shown in Figure
\ref{fig:tcut}, where we have used $1/m_a(\tcut) t_*$ as the x-axis.
This is roughly the inverse of the number of times the axion field
oscillates and it indicates how adiabatic the axion oscillations have
become.  To get closer to 0 in this variable requires tightening the
lattice spacing, linearly in this variable.  We also plot what happens
when we change the lattice spacing, or more precisely, when we vary
$t_*/a$, keeping $m_a(\tcut) a$ fixed.
This tests the same physics, but with added contamination of
lattice-spacing (finite $r_0/t_*$) artifacts at the largest values.
The axion-number estimates from the two string-cutting methods move
towards each other as we postpone the cutting, albeit very slowly.  It
appears that the first/second method converge from above/below, in
which case we can use them as upper and lower systematic-error limits
on the actual axion production.  The sensitivity to the string-cutting
procedure is modest, representing $\sim 15\%$ of the axion
production.  But it will dominate over statistical errors and it
limits our ability to extract a final axion number density.

\begin{figure}[htb]
\centerline{\epsfxsize=0.55\textwidth\epsfbox{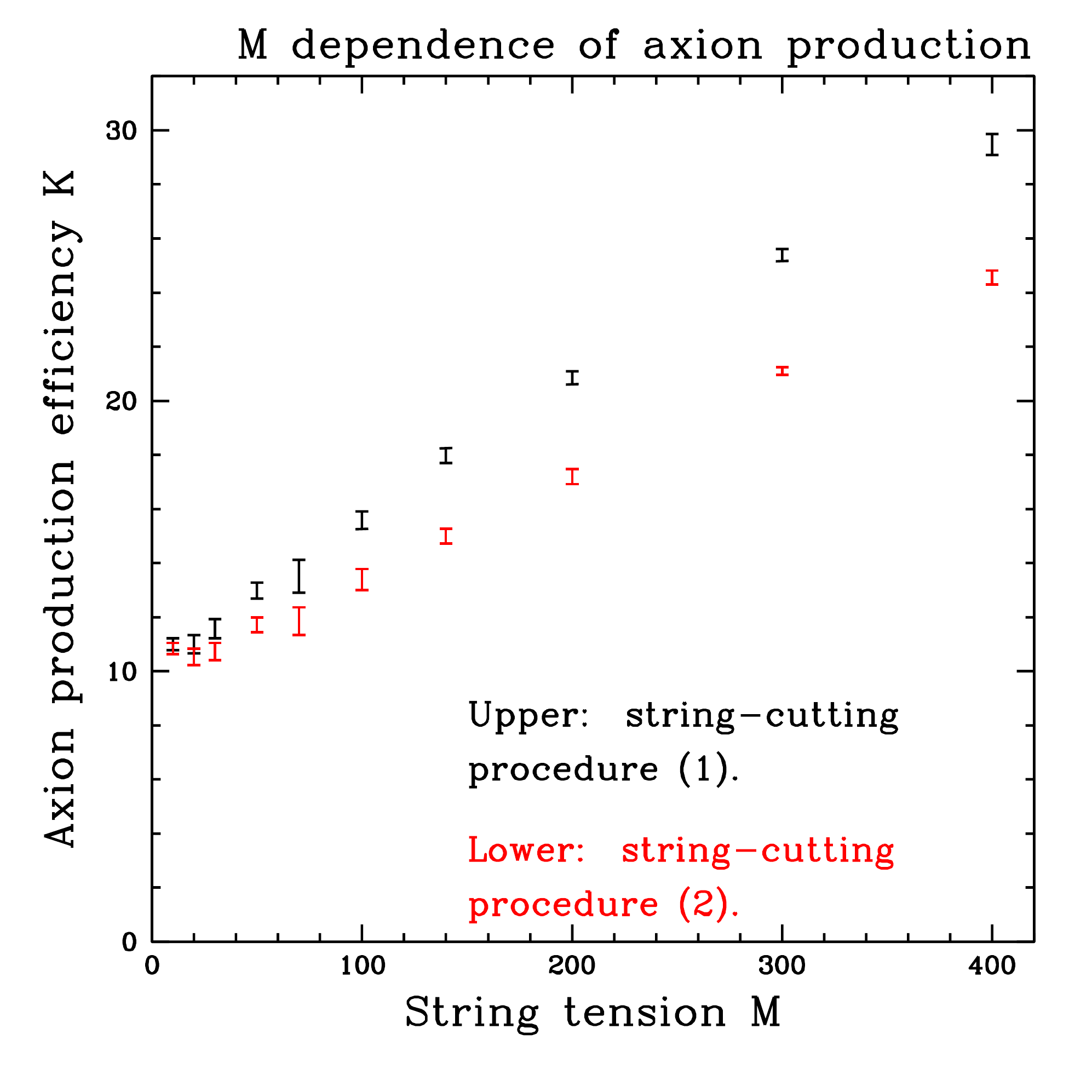}}
  \caption{$M$ dependence of axion number production efficiency $K$.
    Upper (lower) curves are the first (second) string-cutting
    procedure described in the text, and represent upper (lower)
    bounds on the systematic error, due to the longevity of orbiting
    string pairs. \label{fig:money}}
\end{figure}

Finally, we explore the $M$ dependence of the axion number
production.  Figure \ref{fig:money} shows our results, which indicate a
roughly linear rise with $M$ in the axion production efficiency.  For
the physically relevant value $M \simeq 200$, we find $K \in
[17,20]$.  The inclusion of large string tension has roughly doubled
the axion production efficiency, relative to the results from
fields-only simulations.

We postpone an interpretation of these results to the discussion
section.

\section{Extension to 3+1 dimensions}

Suppose that we have an algorithm for evolving a Nambu-Goto string in
3+1 dimensions.  Several such algorithms already exist
\cite{Albrecht:1989mk,Bennett:1989yp,Allen:1990tv,%
Vanchurin:2005yb, Olum:2006ix}.
We will require an algorithm which describes the string as a series of
straight segments (or equivalently as a series of neighboring points,
in which case we take the segments to be the line segments connecting
these points).  Our goal is to present an algorithm for implementing
the interactions between these string segments and the axion field.

Consider a string with affine parameter $\sigma$.  The location and
velocity of the string are $y_i(\sigma,t)$ and
$v_i(\sigma,t) = \partial_t y_i(\sigma,t)$ with $v^0=1$,
$v_i \cdot y_i'=0$, where $y_i' \equiv dy_i/d\sigma$.  The latter is a
gauge choice, that the velocity is at 
right angles to the string's extension.  The action for
the string and the $\tA$ field around it have been nicely discussed by
Dabholkar and Quashnock \cite{Dabholkar:1989ju}.  The string core
should obey a Nambu-Goto action plus an interaction
with the $\tA$ field as follows.  If we define the dual Kalb-Ramond
\cite{Kalb:1974yc,Vilenkin:1986ku} field strength
\be
H_{\mu\nu\alpha} = + f_a \epsilon_{\mu\nu\alpha\beta} \partial^\beta \tA
\label{HKR}
\ee
then $H$ feels a current from the string,
\be
\label{EOMKR}
\partial_\mu H^{\mu\alpha\beta} = j^{\alpha\beta} \,,
\qquad
j_{\mu\nu}(x) = -2\pi f_a \int d\sigma \delta^3(x-y(\sigma))
( v_\mu y_\nu' - y_\mu' v_\nu ) \,,
\ee
with $y' = dy/d\sigma$.  That is, just as in the 2+1 dimensional case,
the string is responsible for a $\delta$-function contribution to the
curl in $\partial_\mu \tA$ which lies along the string's extension.
We will again smear out the exact location of the Kalb-Ramond current,
\be
j_{\mu\nu}(x) \to -\frac{f_a}{2} \int d\sigma g_3(x-y_i(\sigma))
( v_\mu y_\nu' - y_\mu' v_\nu ) \,,
\label{EOMKRsmear}
\ee
where $\int r^2 g_3(r) dr = 1$ so $\int d^3 x\: g_3(x) = 4\pi$.  Again
we define $f_3(r)=\int^\infty_r g_3(r_1) r_1^2 dr_1$, which varies
from 1 at small $r$ to 0 for $r>r_0$, and then revert to the
$\tA$-field description.  The derivative of the $\tA$ field should be
modified to a covariant derivative
\bea
\label{DtAagain}
\LL_{\tA} &=& -\frac{f_a^2}{2} D_\mu \tA D^\mu \tA \,, \qquad
D_\mu \tA(x) = \partial_\mu \tA(x) - A_\mu(x) \,,
\\
A^\mu(x) &=& \int d\sigma \epsilon^{\mu\nu\alpha\beta}
\frac{v_\nu r_\alpha y_\beta'}{2r^3}  f_3(r)
\eea
with $r_i=x_i-y_i(\sigma)$.  The force per unit $\sigma$ acting on the
string is
\be
dF^\mu(y) = \frac{f_a^2}{2} \int d^3 x \: g_3(r) \epsilon^{\mu\nu\alpha\beta}
 v_\nu y'_\beta D_\alpha \tA
\label{F3D} 
\ee
which must be incorporated into the string's equation of motion.
The usual equation of motion \cite{Turok:1984db}, defining
\be
\label{epsdef}
\varepsilon \equiv \mu \sqrt{\frac{(y')^2}{1-(\partial_t y)^2}} \,,
\ee
is modified to
\be
\label{stringEOM3}
(H+\partial_t) ( \varepsilon \partial_t y_i)
= \partial_\sigma ( \mu^2 \varepsilon^{-1} y_i') + dF_i \,.
\ee
Here $\mu=\pi f_a^2 \kappa$ is the string tension, which plays the
role of $M$ in the 2+1 dimensional theory.  Again \Eq{F3D}
automatically ensures that $\partial_t \varepsilon = dF^0 = v_i dF_i$,
and \Eq{stringEOM3} becomes
\be
\label{stringEOMv2}
\partial_t^2 y_i = -H \partial_t y_i +
\frac{\mu^2}{\varepsilon} \partial_\sigma ( \varepsilon^{-1} y_i')
+ \frac{(\delta_{ij} - v_i v_j)}{\varepsilon} dF_j \,.
\ee
The $(\delta_{ij}-v_i v_j)$ term is the usual relativistic reduction
of the acceleration along the direction of motion.

The special case in which $\tA$ is $z$-independent and all strings
stretch strictly in the $z$-direction is equivalent to the 2+1D
smeared-charge theory we previously presented, after identifying
$g(r) = \half \int dz g_3(\sqrt{r^2+z^2})$.  

The lattice implementation is as follows.  The string must be
considered as a series of short segments.  The $b$'th segment has a
basepoint $y^i_b$ and an extent $s_b^i=y^i_{b+1}-y^i_b$, which plays the
role of $y'$.  \Eq{F3D} for a string segment is found by summing over all
links close to a segment, using $r$ in $g_3(r)$ as the distance from
the center of the link to the center of the string segment and
replacing $y_b'$ with $s_b$.  And
\be
\label{A3Dlatt}
A^\mu(x) = \sum_b f_3(r) \frac{\phi(x,y_b,s_b,\mu)}{2} \,.
\ee
Here $r$ is evaluated as the distance between the midpoint of the
string segment and the midpoint of the lattice link for $\mu=i$, and
as the distance from the lattice point to the middle of the string segment
halfway between time $t$ and $t+\delta a$ for the $\mu=0$ case.  And
$\phi$ is a solid angle which is determined as follows.  For $\mu=j$
it is the solid angle swept out by the string segment as one changes
perspective by sliding along the lattice link.
Equivalently, it is the solid angle, as seen from the base point $x$ of
the link, of the parallelogram with corners $\vec y_b$, $\vec y_{b+1}$,
$\vec y_{b+1}-a\hat{j}$, and $\vec y_b-a\hat{j}$.  For $\mu=0$ it is
the solid angle swept out by the motion of the string segment from
time $t$ to time $t+\delta a$, as seen from the lattice site.

It is clear that the lattice
part of the update can be accomplished without much more difficulty
than in the 2+1D case (though of course 3+1D simulations will be much
more expensive numerically).  It is not clear to us how best to
implement \Eq{stringEOMv2}, but we believe that it should be possible
to modify known Nambu-Goto algorithms to incorporate the force term.
Nor is it clear to us, at present, whether it will be necessary
to include explicit short-range inter-string interactions and
radiation-reaction effects like the ones we used in the 2+1D theory.
It is also necessary to keep track of when string segments intersect,
since global strings are generally expected to intercommute
\cite{Shellard:1987bv}.

\section{Discussion}

Axions present a well motivated dark matter candidate.  With the
additional assumption that PQ symmetry is restored in the early
Universe, the model \textsl{should} be predictive in the sense that
there should be a clean relation between the axion mass and the axion
dark matter abundance.

The main stumbling block to finding this relation is the efficiency of
axion production.  This is hard to determine because it depends on the
behavior of axionic strings, and the string dynamics are sensitive to
the string tension, which varies logarithmically with the ratio
$f_a/H$.  In nature (assuming axions exist) the ratio is
$\sim 10^{30}$, while in fields-only numerical simulations it is
$\leq 10^3$.

We have presented a new algorithm for solving this problem, by
treating the string cores as additional explicit objects in a
simulation of the axion field.  We have presented, implemented, and
studied this method in 2+1 dimensional space, and we have shown
how it could be extended to 3+1 dimensions.

The axionic string networks in 2+1 dimensions are sensitive to the
logarithm $\kappa \equiv \ln(f_a/H)$, with the density of strings
rising roughly as $\kappa^{3/2}$ and the axion production increasing
linearly with $\kappa$.  Unfortunately, the physics of the strings
which leads to this behavior does not look very much like the string
network dynamics we would expect in 3+1 dimensions.  The strings bind
into long-lived orbital pairs, which are actually longer-lived and
more numerous when the potential ``tilts'' than for the case with no
explicit $U(1)$ symmetry breaking.  The longevity of these systems is
partly because it is difficult to radiate massive axion excitations,
and partly because they are nonrelativistic.

In 3+1 dimensions, we expect that increasing $\kappa$ will lead to a
denser string network which will produce more loops.  Also, since
strings radiate less efficiently, the loops can be longer-lived.  But
the motion of a string in 3+1 dimensions is generically relativistic,
due to string tension effects.  So 3+1D string loops should radiate much
more effectively than the bound string pairs of the 2+1D system.  As
the explicit symmetry breaking becomes important, the long strings
should strongly attract each other and annihilate or fragment into
loops.  So at, say, $t=3t_*$, we expect 3+1D simulations to contain
axions and small string loops.

It is possible that these loops will be long-lived, as their ability
to radiate away their energy may be sufficiently suppressed because of
the axion mass.  So consider the case where they lose \textsl{no}
energy to radiation of axions.  In this case their energy density
dilutes under Hubble expansion as $a^{-3}$, like matter.  But the
axions dilute like $a^{-3} m_a$, and
$m_a \propto T^{-7/2} \propto a^{7/2}$ so long as the axion mass
remains temperature dependent.  Then the energy stored in string loops
dilutes away relative to already-produced axions as $t^{-7/2}$, a much
stronger power than \Eq{eqdilute}.  Relative to our 2+1D simulations,
any long-lived string structures should not play much of a role in 3+1
dimensions.  Radiation from strings might still be important in 3+1D,
even more important than in 2+1D, because of another difference.  In
2+1D, almost all of a string's energy is always lost in the ``final
inspiral'' to very short-wavelength axions with essentially no axion
number.  In 3+1D, cusps, bends, and waves along strings can turn
string tension into axions, and a bigger fraction of the string
network's energy may go into long-wavelength radiation.

For completeness we will update our results on the implied axion mass
from \cite{axion1}, \textsl{assuming} that the axion production
efficiency in 2+1D is the right one for the physical case of 3+1D.
Taking the production efficiency to be $K=19$ (between
the upper and lower estimates for $M=200$ in Figure \ref{fig:money}),
and applying Eq.~(5.1--5.5) of \cite{axion1}, we find
$T_*=1.72$GeV, $f_a=1.6\times 10^{11}$GeV, and $m_a=36\mu$eV.
\textsl{However}, our discussion of the differences between the 2+1D
and 3+1D cases gives us little confidence that the axion production
efficiency in 2+1D is the same as in 3+1D.  It is not even clear to us
whether to expect the 3+1D axion production to be larger or smaller.
What we have learned is that increasing the string tension really does
increase axion production.  But the above results can only serve as a
crude guideline, not a real calculation.

It should be clear that 3+1 dimensional simulations, with strings as
explicit objects which couple to $\tA$ fields, are needed.  We have
presented a nearly-complete algorithm for performing such simulations.
The numerical effort to study the 3+1
dimensional problem will be much ($\sim 10^3\times$) larger than what
was required here, but all of the results in this paper were generated
in a few weeks on a single laptop, so 3+1D studies should be feasible.

\section*{Acknowledgments}

We would like to thank Jeorg Jaeckel, Arthur Hebecker, Edmond Iancu,
and Jean-Paul Blaizot for interesting conversations.
This work was supported by the Natural Science and
Engineering Research Council (NSERC) of Canada.

\appendix

\section{Charge inspiral}
\label{apprad}

In the main text we need to know how efficiently a pair of vortices
lose energy when they become bound to each other.  This is equivalent
to asking how an isolated, bound pair of electric charges in 2+1
dimensional electrodynamics radiates away energy.
This is important both for understanding what to expect when charges
get close together, and for adding explicit radiation when the charges
are very close and we must incorporate radiation explicitly to account
for its suppression due to a form-factor arising from the way we
spread the charge into a ball.  Unfortunately, there is no Larmor
formula for radiated power in 2+1 dimensional electrodynamics, nor is
there an Abraham-Lorentz expression for the radiation-reaction force.
The reason is that Huygens' principle does not apply in 2+1
dimensions; the 4-vector field due to a charge is not determined by
the 4-currents on the backwards light-cone of the field point, but by
all 4-currents inside the light-cone.

We will first
solve for the motion ignoring radiation and relativistic corrections.
We will use this solution to determine the radiated power, and use it
to determine how the charges must inspiral.
For simplicity we will only solve for the case of
a nearly-circular orbit.  We find that the orbital velocity is
constant and the charges follow an exponential spiral.
Throughout, we scale out the irrelevant overall factors of $f_a$.

Consider two charges with $q=\pm 2\pi$ and charge separation $R$, so
they are each $r = R/2$ away
from the common center of mass.  They each have mass $M$ and an
attractive force of strength $2\pi/R$ acts between them.  Therefore,
neglecting radiation reaction, they should follow circular orbits with
velocity and frequency
\be
\label{vorbit}
\frac{\pi}{r} = F = \frac{Mv^2}{r} \quad \Rightarrow \quad
v = \sqrt{\pi/M} \,, \qquad
\omega = v/r = \sqrt{\pi/Mr} \,.
\ee
To find the associated radiation power, we consider the 2+1D theory as
the same as the 
3+1D theory with infinite line charges.  We need to compute the
far-field Li\'enard-Wiechert potential arising from the
current-per-length of each charge.  The charges have opposite
current-per-length, each $2\pi v$, so the $A_\phi$ potential at a
distance $d \gg 1/\omega$ from the charge pair is
\be
\label{Aphi}
A_\phi(d) = \frac{2\times 2\pi v}{4\pi} \Re \int_{-\infty}^\infty
   \frac{d\ell}{\sqrt{d^2+\ell^2}}
   e^{iv\sqrt{d^2+\ell^2}/r} e^{-i\omega t}
\ee
where $\ell$ is the distance along the (fictitious) 3'rd direction.
The phase factor here is just the retarded phase
$e^{-i\omega(t-D)}$ with the source distance $D=\sqrt{d^2+\ell^2}$.
Expanding in large $d$, we find
\be
\label{Aphi2}
A_\phi(d) = v \Re \int_{-\infty}^{\infty} \frac{d\ell}{d}
e^{-i\omega t +ivd/r} e^{\frac{iv\ell^2}{2dr}}
= \frac{v}{d} \Re e^{i\ldots} \sqrt{\frac{2\pi dr}{v}}
= \cos(\ldots) \sqrt{\frac{2\pi vr}{d}} \,.
\ee
The wave number is $k=\omega=v/r$ and $B=kA$.  The power carried away
by the wave is
\be
\label{power1}
P = 2\pi d \frac{E^2 + B^2}{2} = 2\pi d B^2
= \pi d k^2 A_{\phi,\mathrm{peak}}^2
\ee
where the lost 2 is the average of the $\cos^2$.  Substituting,
\be
\label{power2}
P = \frac{\pi d v^2}{r^2} \: \frac{2\pi vr}{d}
 = \frac{2\pi^2 v^3}{r} \,.
\ee

This power is to be compared to the potential
\be
\label{V1}
V = \frac{q^2}{2\pi} \ln(2r) = 2\pi \ln(2r) \,.
\ee
Here $2r=R$ is the separation of the charges.  The value of $r$ must
evolve such that the power is the time derivative of the potential:
\be
\label{match1}
\frac{2\pi^2 v^3}{r} = \frac{dV}{dt} = \frac{2\pi}{r} \frac{dr}{dt}
\,.
\ee
We can then find $v_r = dr/dt$ and the time for the inspiral to
complete:
\be
\label{inspiral}
\frac{dr}{dt} = -\pi v^3 \,, \qquad
t_{\mathrm{inspiral}} = \frac{r}{dr/dt} = \frac{r}{\pi v^3}
= \frac{r M^{3/2}}{\pi^{5/2}} = \frac{R M^{3/2}}{2\pi^{5/2}}  \,.
\ee
The angle of the inspiral is fixed,
\be
\label{inangle}
\sin \theta_{\mathrm{inspiral}} \equiv \frac{-v_r}{v_\phi}
 = \frac{\pi v^3}{v} = \frac{\pi^2}{M} = \frac{\pi}{\kappa} \,.
\ee
We have tested this description against the actual behavior of our
code, for pairs of charges alone in a large box at separations
$r > r_0$ but $r \ll Na$ the lattice size, and found good agreement.

Note that the inspiral process gives rise to an equal amount of energy
in each logarithmic interval in frequency.  Almost all of the charges'
energy is released into extremely short wavelength fields, which carry
very little axion number per unit energy.  Therefore, once the charge
separation has become smaller than the resolution scale of a
simulation, it is safe to ignore the axion number produced by the
subsequent inspiral, even though the energy release represents most of
the energy present in the strings.

We have not solved for the case of a highly noncircular orbit, but we
performed numerical experiments which suggest that the
radiation-reaction force near the charges' closest passage is
\be
F_{\mathrm{rad.react.}} \simeq -\frac{\pi^3}{Mr} \hat{v} \,,
\label{Fradreact}
\ee
which for the case of circular motion coincides with
$F=-\pi^2 |v|\vec{v}/r$ as one would guess from \Eq{power2}
(remembering that half the power is radiated from each charge).  When
the charges are far from closest approach, we find that the reactive
force falls below the above estimate.

Generally the radiation-reaction force arises automatically from the
coupling between the charges and the electromagnetic ($\tA$) field.
But in our lattice implementation we replace point charges with
finite-sized charge balls.  Any radiation with wavelength satisfying
$kr_0 \leq 1$ is suppressed, since the current $Qv$
is replaced by $v\int_y \rho(y) e^{i\vec k\cdot \vec y}$.  Phase
cancellation in this integral results in a form-factor suppressing the
coupling to the radiative part of the $\tA$ field.  Therefore
we need to add a radiation-reactive force explicitly whenever the
charges get so close that this effect suppresses their radiation.
Failing to do so results in charges which stop inspiraling once they
get sufficiently close.  It is not necessary to include this effect
with high precision, since it primarily affects strings which are
about to annihilate; but we should include it to ensure that the
annihilation really occurs.  We do so by applying a reactive force
equal to \Eq{Fradreact} times $f(2vR/r_0)$, a crude estimate of the
squared form factor.  We could estimate the form-factor more
accurately, but we have not done so because \Eq{Fradreact} is already
only an estimate.

The efficiency of radiation changes fundamentally if the axion mass is
turned on.  In this case, for an axion mass $m_a$, the radiation is
suppressed whenever $\omega < m_a$, which for a circular orbit is when
$r < v/m_a$ or $m_a r < v$.
This is different from $m_a r < 1$, the criterion that the potential
is significantly changed from the log form.  For large $M$,
radiation is suppressed even for charges far too close together for
any ``domain wall''-like behavior in their interaction energy.
We have not found a good way to compute the radiated power in this
case, but numerical experiments show that the suppression is severe
and radiative energy loss essentially does not happen.  We believe
that this is a peculiarity of the 2+1D approximation which should be
much less severe in 3+1D, where the string velocities will always be
relativistic.

\section{Checks and tests}
\label{checks}

Our numerical implementation contains several parameters, such as
$\delta$, $r_0/a$, $\rmin$, and the initial time $t_i$ and
amount of smearing used in the initial conditions.  We have to ensure
that there is some range for each parameter where we get
continuum-like behavior, and determine at what time scale we achieve
the scaling solution for the defect network.  Our goal is results for
the final axion number produced with
few-percent systematic sensitivity to our implementation.

Let us start with $\delta$ and $\rmin$, the temporal step and the
separation at which strings are taken to annihilate.  Physically we
want the limit where both go to zero, but numerically this is
impractical.  The bare minimum value for $\delta$ is set by the
smaller of the Courant condition for the lattice field update,
$\delta^2 < 1/2$, and the Courant condition%
\footnote{%
  By the Courant condition we mean the $\delta$-value beyond which
  some lattice mode or some pairing of strings will undergo
  oscillatory growth rather than stable evolution.  Violating the
  condition for strings makes it difficult or impossible for pairs to
  annihilate; violating it for the fields leads to exponentially
  growing energy and fluctuations.}
for the last steps of the
inspiral of annihilating strings, $\delta^2 < M\rmin^2/\pi$.  Also,
our radiation-reaction method ceases to be energy-reducing when the
separation approaches $\rmin$ unless $\delta^2 < M\rmin^2/(2\pi)$.
It is less clear what the requirements on $\rmin$ are. The larger the
value of $\rmin$, the more violent the process of removing two strings 
becomes; we also lose some radiation from the final inspiral if
$\rmin > v r_0$.

We investigate $\rmin$ first.  We fix $\delta=1/20$ to be small, and
consider $M=200$ (corresponding to $\kappa=64$, around the physical
value), 10 passes of checkerboard field smearing in the initial
conditions, $t_i=0$, and $r_0=4a$, which we will later see is more than
sufficient.  We then consider the density of strings and the energy
components in the $\tA$ field (without any tilt to the potential) for
several values of $\rmin$.

\begin{figure}
  \epsfxsize=0.4\textwidth\epsfbox{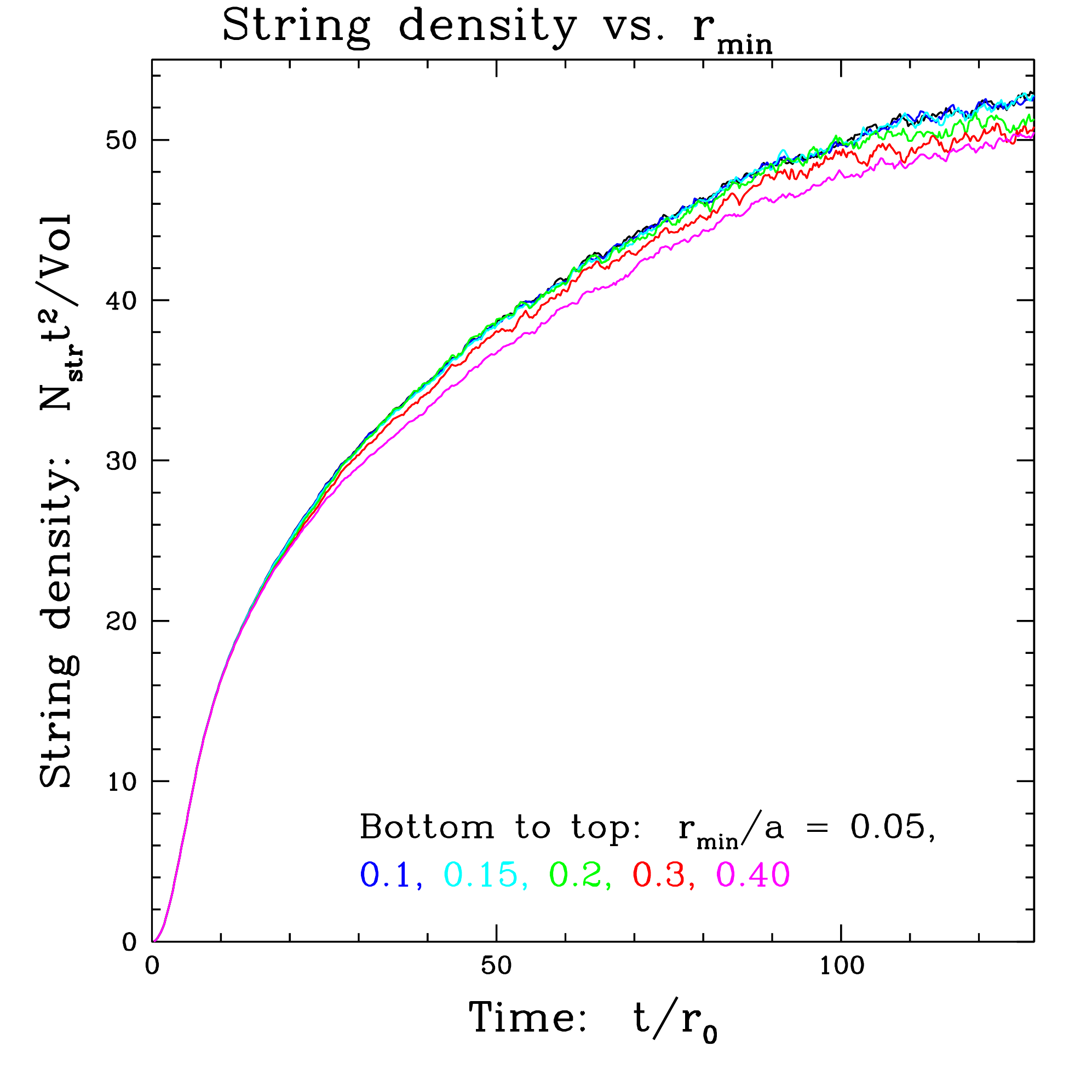}
  \hfill
  \epsfxsize=0.4\textwidth\epsfbox{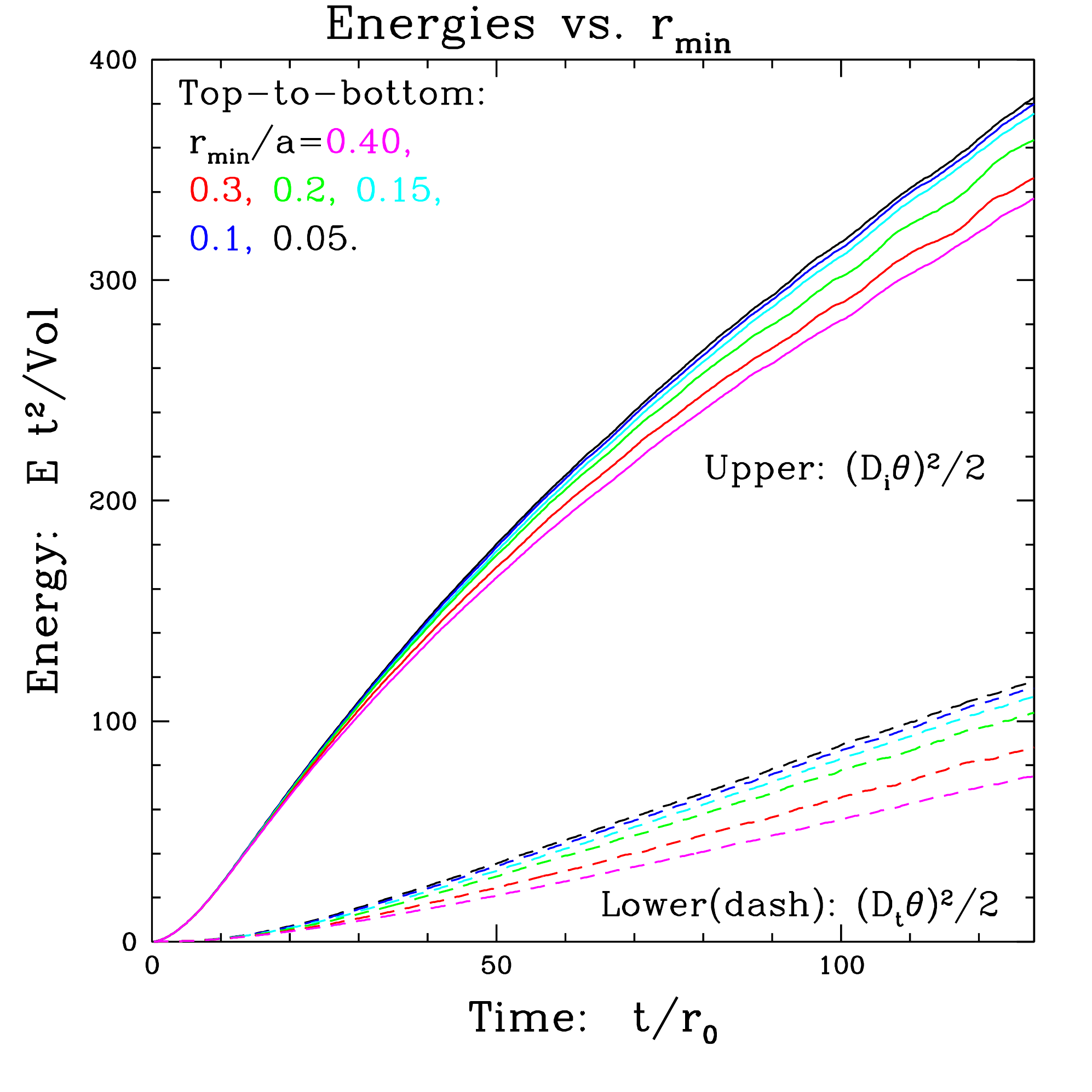}
  \caption{\label{thresh_fig} The dependence of vortex number (left)
    and scaled energy densities (right) on the cutoff separation where
    strings are taken to annihilate.  The largest values show clear
    deviation, but for $\rmin=0.15 a$ and smaller the differences are
    negligible. }
\end{figure}

The results are shown in Figure \ref{thresh_fig}.  We used the same
random number seed for each $\rmin$ value, so the fluctuations in
different curves are highly correlated and it is easier to see small
differences over statistical noise.  We plot the number of vortices in
the $\tA$ field, rather than the number of strings, because once the
strings get closer than $\sim a/2$ the associated vortices disappear,
but different $\rmin$ values will consider the string pair to still
exist for different lengths of time as they inspiral.  Therefore the
number of vortices avoids a trivial reason for the string counts to
differ.  We also compare the energy content in $(D_i \tA)^2$ (gradient
energy) and $(D_0 \tA)^2$ (kinetic energy).  It is clear from the
figure that too
large a value of $\rmin$ causes strings to annihilate too soon,
lowering the string density and also reducing the amount of field
energy radiated before the strings annihilate.
On the other hand, smaller values clearly approach a good
limit.  We will use $\rmin=0.1$, since smaller values increase
numerical cost without any clear benefit.

\begin{figure}
  \epsfxsize=0.4\textwidth\epsfbox{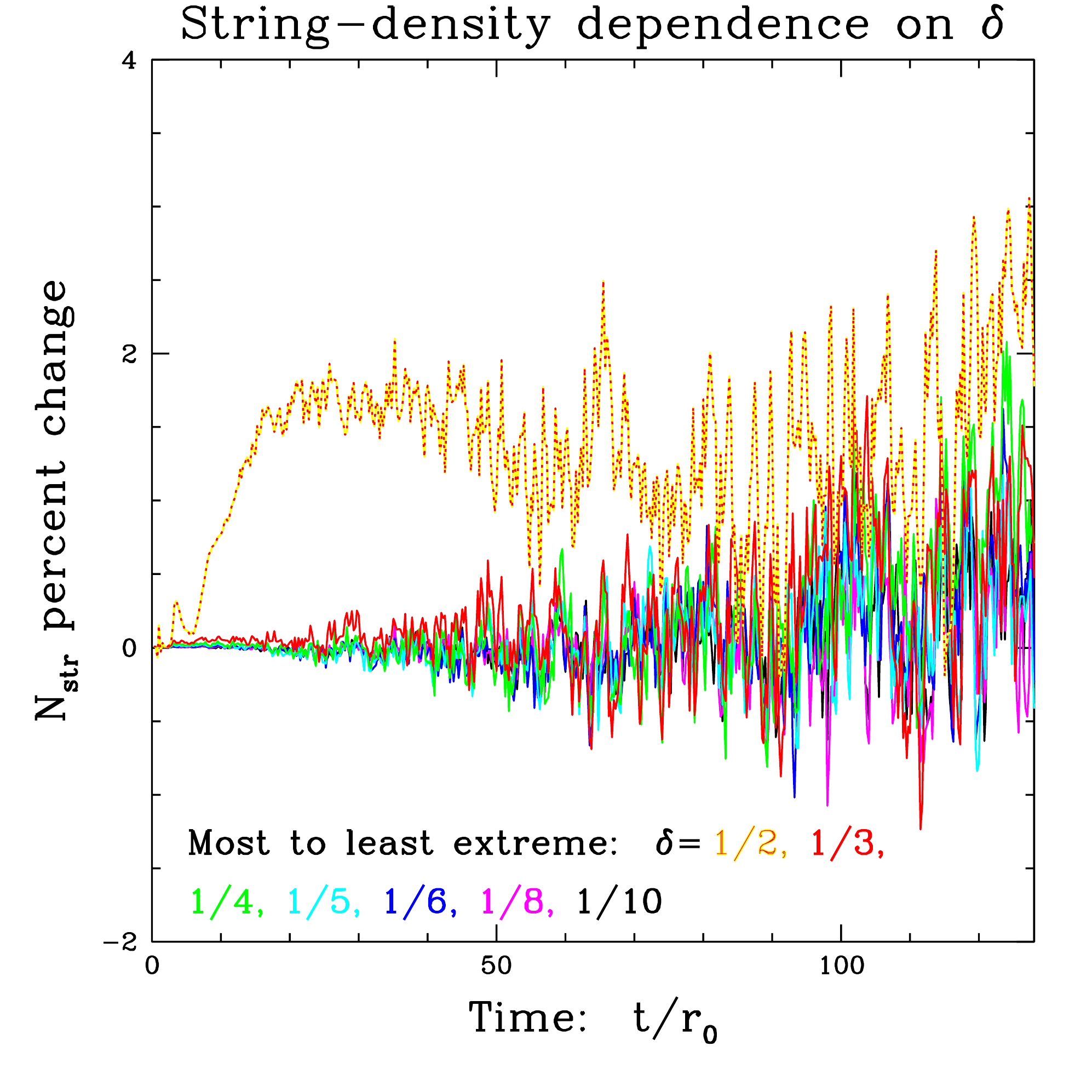}
  \hfill
  \epsfxsize=0.4\textwidth\epsfbox{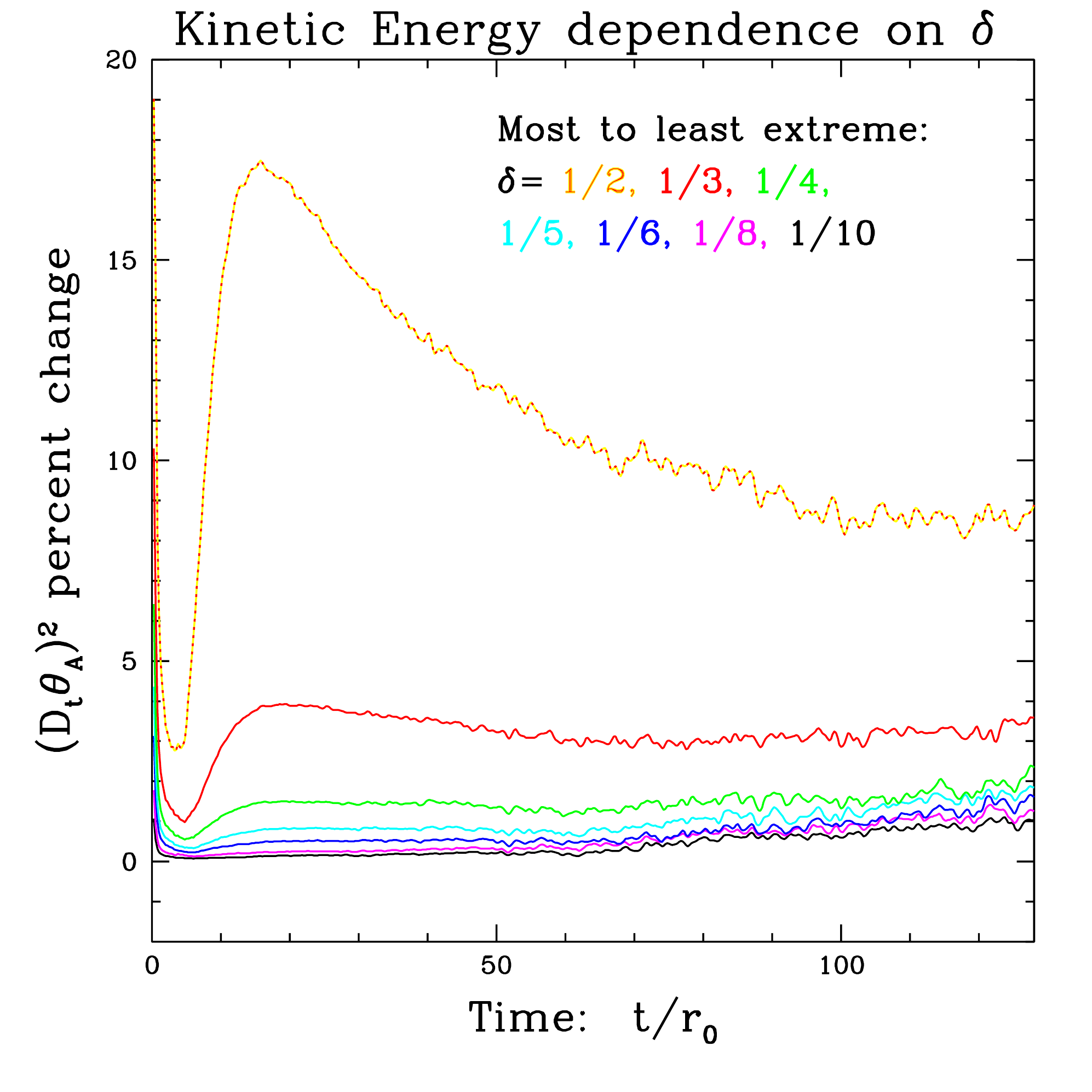}
  \caption{\label{deltafig} The dependence of string number (left)
    and scaled kinetic energy density (right) on the temporal spacing
    $\delta$, shown as a percent change for identical initial
    conditions between the results with a given $\delta$ and the
    results with $\delta=1/20$.  Both are within 1\% for $\delta=1/6$
    or smaller.  Gradient energy shows less sensitivity than kinetic
    energy.}
\end{figure}

Fixing to $\rmin=0.1$, we now consider $\delta$, the temporal spacing.
The results are shown in Figure \ref{deltafig}.  Except for the
coarsest temporal spacings, the value of $\delta$ is strikingly
unimportant.  This is good, as numerical cost scales as $1/\delta$.
We conservatively choose $\delta = 1/6$ for all other studies, which
should keep time-step errors below 1\%.

\begin{figure}[thb]
  \epsfxsize=0.5\textwidth\epsfbox{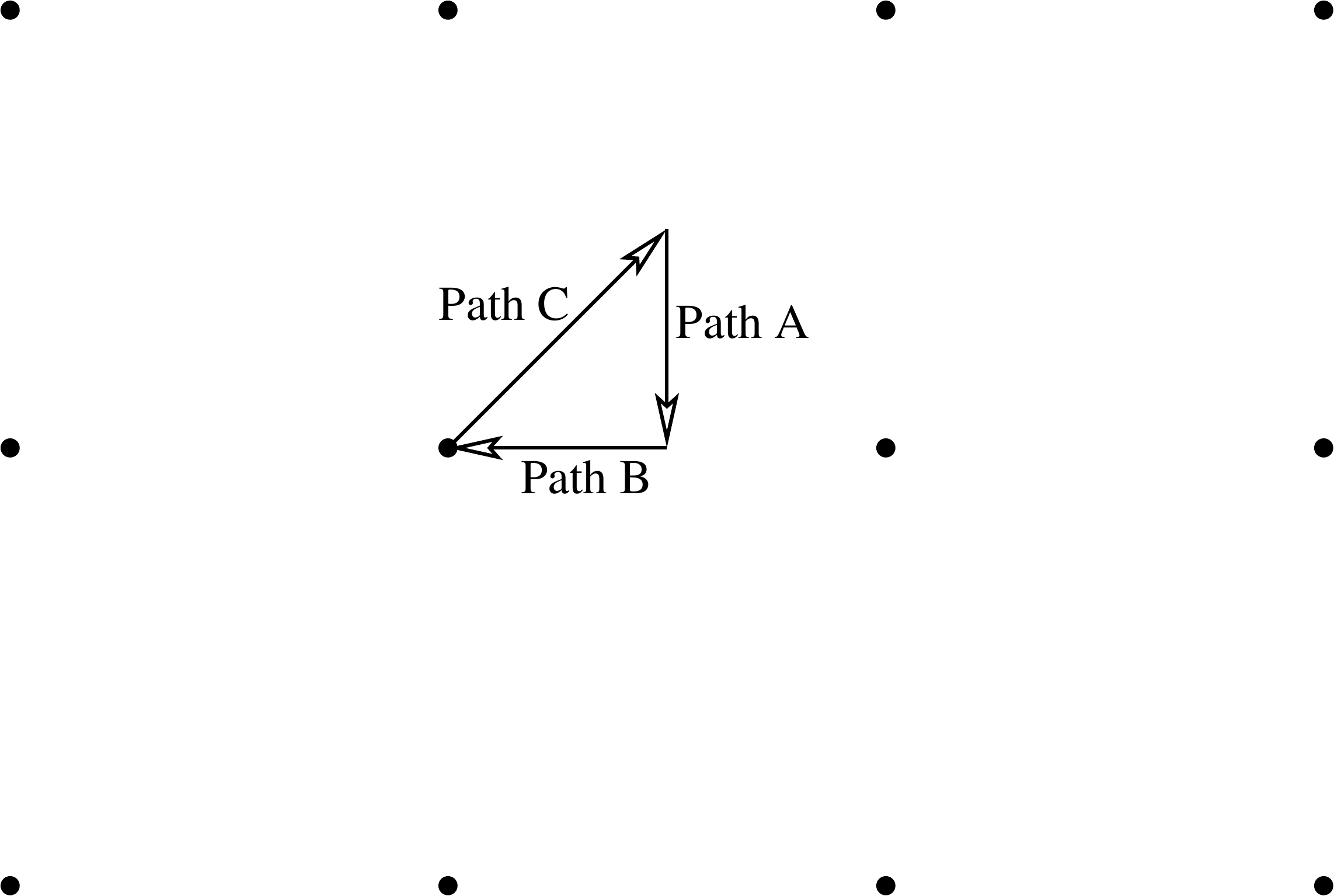}
  \hfill
  \epsfxsize=0.44\textwidth\epsfbox{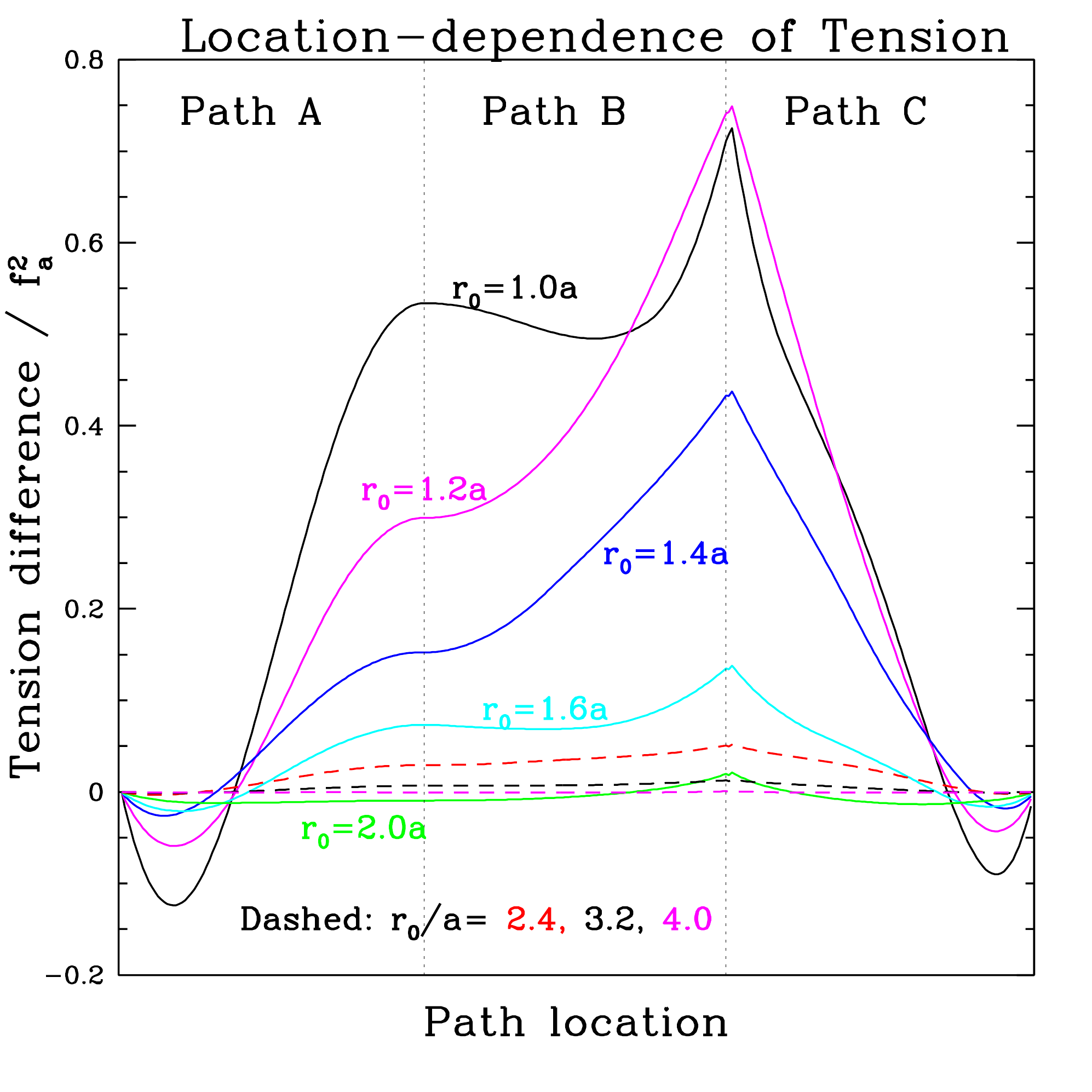}
  \caption{\label{r0fig}
    Left:  a path through a lattice cell, in three steps.  Right:  how
    the string tension (the energy in the $\tA$ field gradients)
    varies as the string is moved along the path, for several values
    of the charge-smearing parameter $r_0$.  Small values $r_0<1.6 a$
    give rise to strong position dependence in the string's energy.}
\end{figure}

Next consider the string core radius $r_0$.  For $r_0/a$ too small the
strings do not have a smooth interaction with the lattice $\tA$
fields.  The gradient energy associated with a string will have a
short-distance contribution which is not translation-invariant, but
depends on where the string sits with respect to the lattice.  We can
investigate this directly by considering lattices with two strings
exactly halfway across the lattice from each other.  We move around
their exact location with respect to the lattice sites, evolving the
lattice fields dissipatively to find the minimal-energy configuration
for a given string location.  The result is shown in Figure
\ref{r0fig}, which shows how the energy varies as the string is moved
along a path through the lattice.  Values $r_0/a \sim 1$ show strong
position dependence, which can interfere with the string's dynamics.
In the units of the plot, a typical string kinetic energy is 2; so the
energy variation through the lattice can trap slower-moving strings so
they stick near the center of a lattice cell, rather than moving
smoothly.  For larger values $r_0 \geq 3$, the energy is a very weak
function of the exact location on the lattice.  For $r_0=4a$ the
variation is invisible in the figure.  For $r_0=2a$ it is accidentally
small for this path.

\begin{figure}[htb]
  \epsfxsize=0.45\textwidth\epsfbox{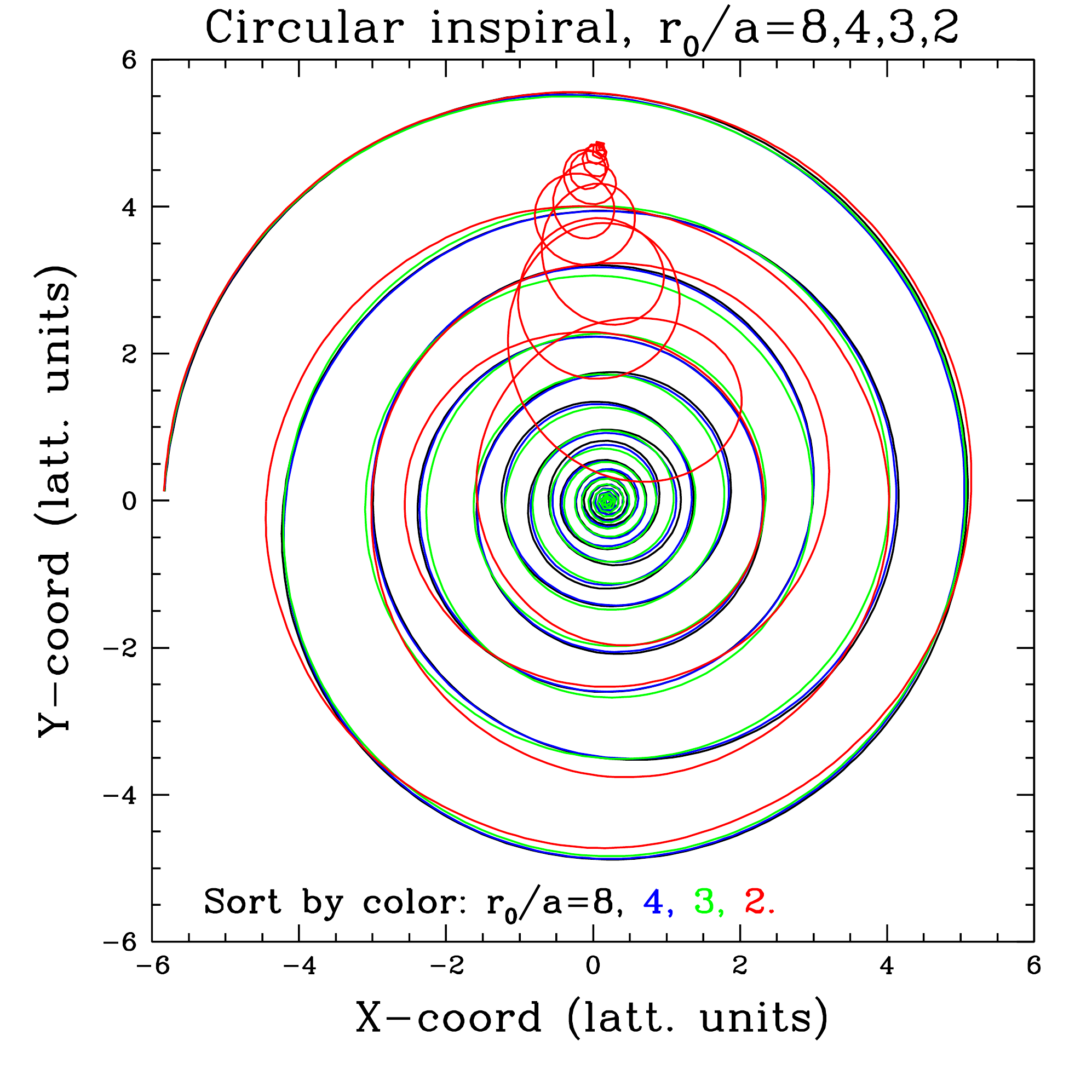}
  \hfill
  \epsfxsize=0.45\textwidth\epsfbox{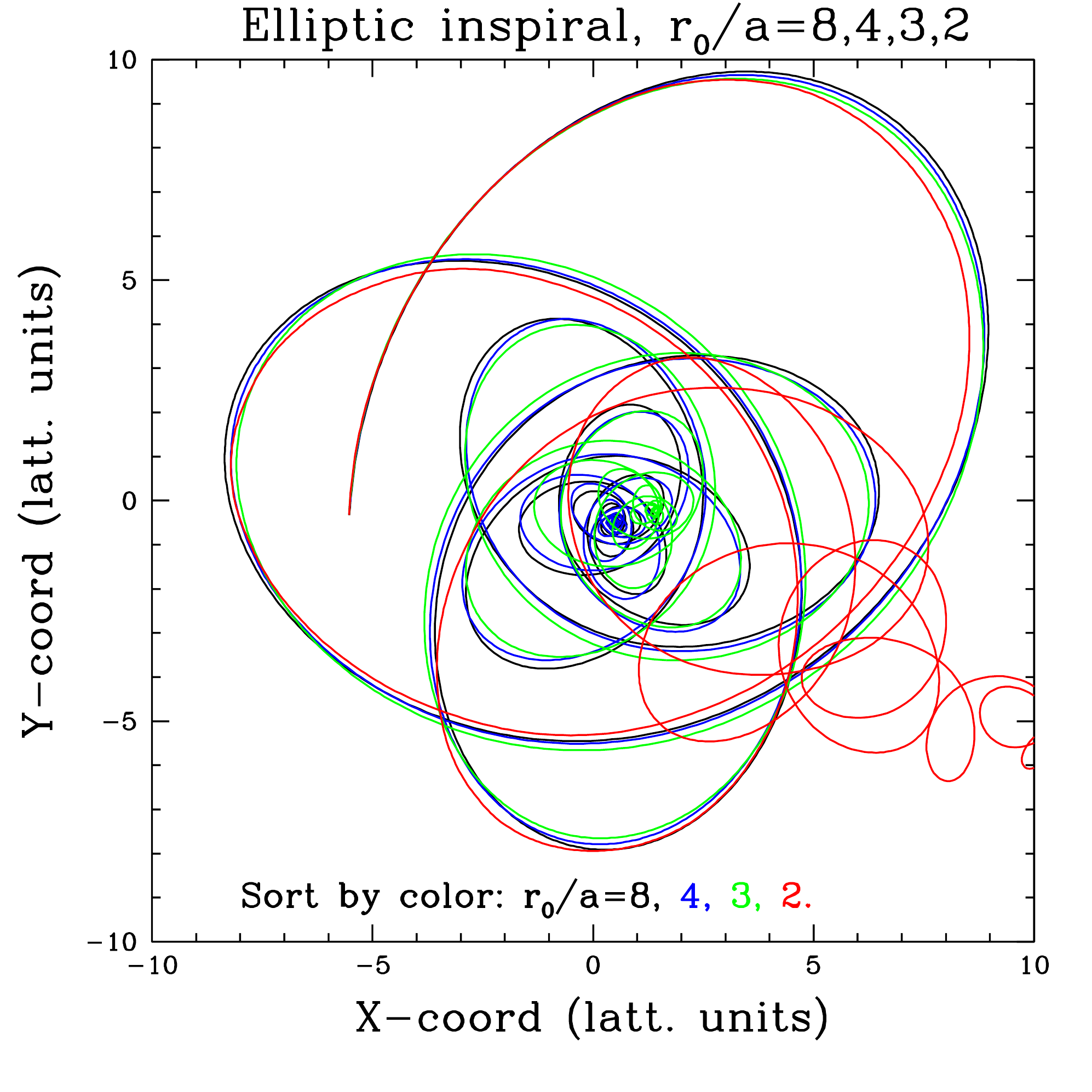}
  \caption{\label{spiralfig}
    The path of one from a pair of orbiting, inspiraling strings,
    initially in a circular orbit (left) or an elliptical orbit
    (right).  The different curves are for different choices of $r_0$,
    the charge-smearing radius.  The values $r_0/a=8$ (black), $4$
    (blue), and $3$ (green) are in good agreement; however, for $2$ (red)
    the charges bounce off of inhomogeneities in their potential with
    respect to lattice position, and the center of mass of the system
    takes on a net motion.}
\end{figure}

As another probe of the impact of $r_0$, we evolve a lattice with only
two strings of opposite charge, initially placed in a circular or
elliptical orbit.  This is a nice example of how orbital inspiral
occurs, and a test of our short-distance force and radiation
additions.  It also lets us see how $r_0$ affects string dynamics.
Figure \ref{spiralfig} shows the coordinate-space track of one from a
pair of inspiraling strings, evolved using several values of $r_0$.
\noindent
The figure shows that, for the smallest choice of $r_0=2a$, the
charges interact with the lattice in a way which occasionally,
abruptly, gives the system a net center-of-mass motion.  This is
clearly an artifact.  Based on the figure, we consider $r_0=3a$ to be
the minimal acceptable value.  Since the numerical cost to update the
strings rises as $r_0^2$ and since larger values make it more
difficult to achieve the small lattice-spacing limit, we will fix to
this value in the following.  The figure also illustrates that
elliptical orbits rapidly precess, which is expected for a $1/r$ force
law or $\ln(1/r)$ potential.

\begin{figure}[htb]
  \centerline{\epsfxsize=0.45\textwidth\epsfbox{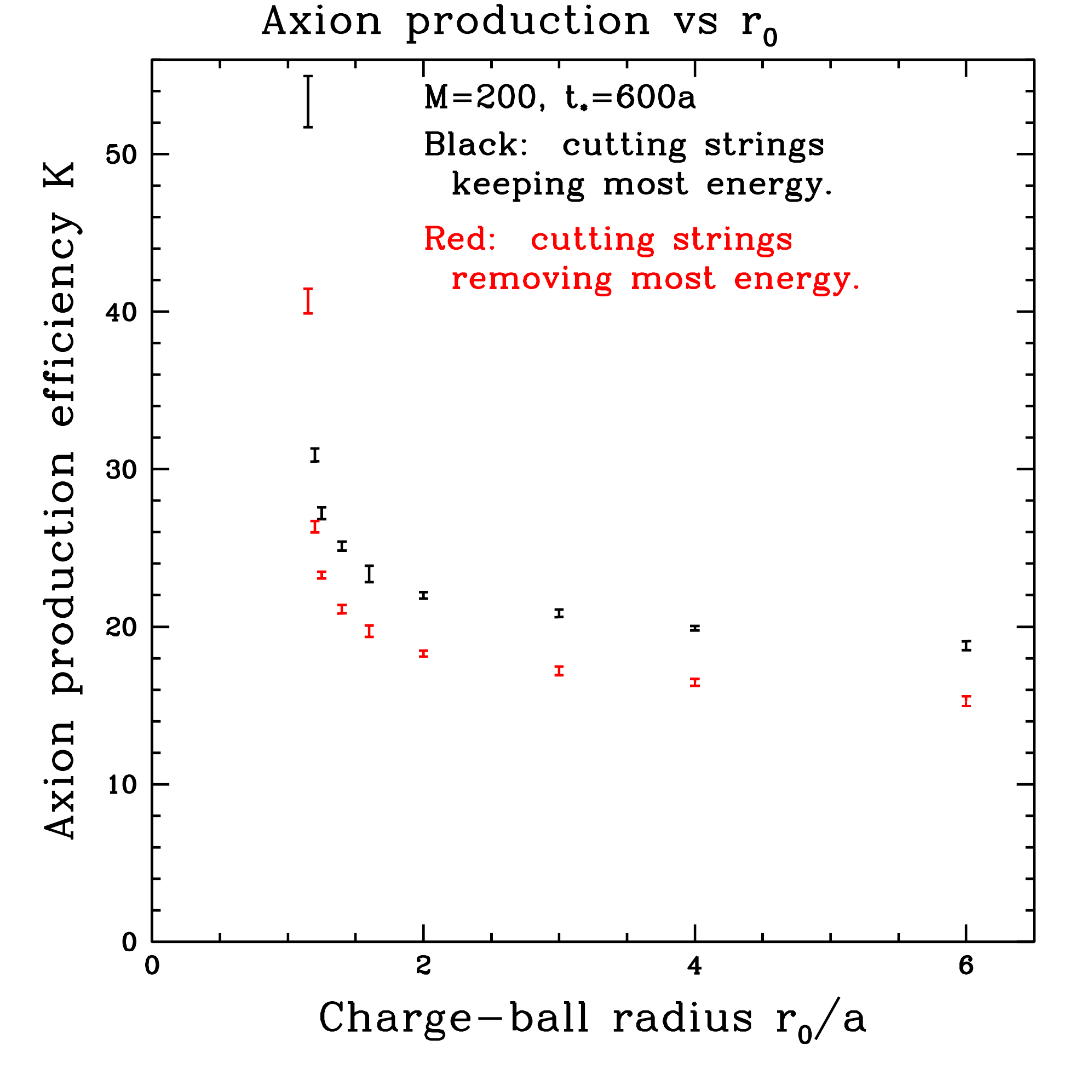}}
    \caption{\label{fig:Kr0}
      Dependence of axion production on $r_0$ choice.  For small $r_0$
      the strings get stuck and are overdense, and axion production is
      large.  For $r_0>2$ there is a plateau, though the production
      falls slowly with increasing $r_0$.}
\end{figure}

As another cross-check on the role of $r_0$, we also compute the axion
number produced, for the $M=200$ case, as a function of the $r_0$
value chosen.  The result is plotted in Figure \ref{fig:Kr0}.  The
procedure is as described in the main text, and as in the main text,
we try two different ways of addressing the strings which persist
until late in the simulation: removing them but leaving most of the
energy in the domain walls connecting them, or removing them and
removing most energy in the domain walls.  These are indicated with
black and red points respectively.  The figure shows that, below
$r_0=2a$, the axion production rises steeply.  For larger values it
falls gradually with increasing $r_0$.  Presumably this arises from
effects where the charged balls are not much smaller than the scale of
structure, or because large charged balls do not feel the right force
strength when the domain walls become thin.  If we make a linear
extrapolation of the $r_0\geq 3a$ data to $r_0=0$, we find about a
10\% increase in the axion production efficiency $K$.  We will treat
this as a systematic error in the main text.

Finally, there is the amount of smearing we perform initially, which
sets the initial density of strings, and the starting time $t_i$.
These parameters can only affect how quickly
the network approaches scaling; at sufficiently late times we should
converge onto the same statistical ensemble of networks, possibly
modulo short-range axionic fluctuations which carry little axion
number.  To study this issue we performed a series of simulations of
string networks, using $t_i=0$ with various numbers of initial
checkerboard smearings and using 2 initial checkerboard steps with
various values of $t_i$.

\begin{figure}
  \centerline{\epsfxsize=0.45\textwidth\epsfbox{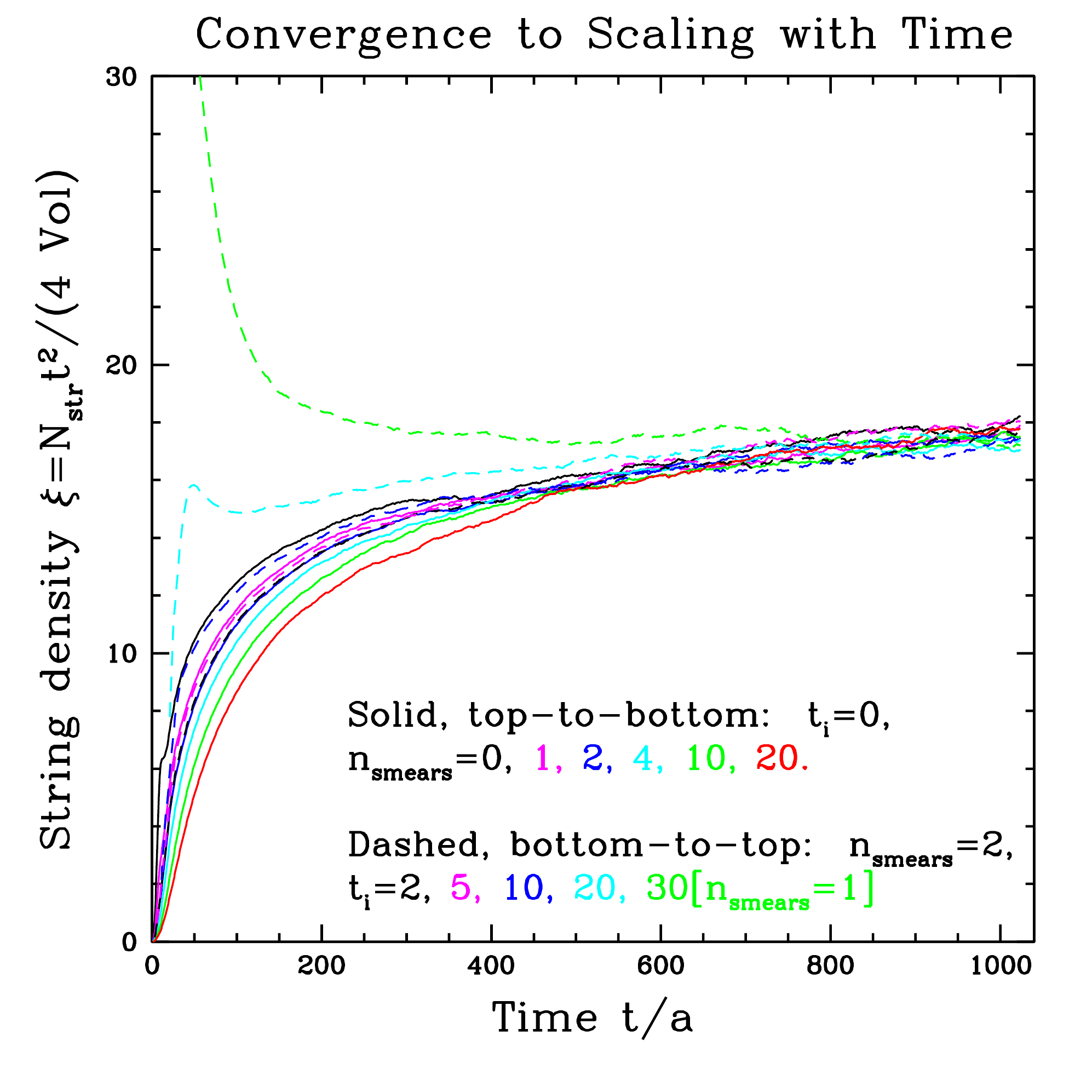}}
  \caption{\label{smearfig}
    Dependence of scaled string density
    on the initial string density, determined
    by the number of smearing steps and the initial time $t_i$. }
\end{figure}

Figure \ref{smearfig} shows how the string density, scaled by a factor
of $t^2/4$ to coincide with the scaling density $\xi$ as defined in
the literature%
\footnote{$\xi$ is defined as the string energy density, scaled by the
  string tension and the system age to form a dimensionless quantity.
  The factor of 4 is because it is conventional to use time and not
  confrmal time; in the radiation era there is a factor of 2 in the
  inter-relation.  Since our strings are nonrelativistic, the number
  of strings is essentially the same as the string energy divided by
  the tension.},
depends on the number of smearing steps and on the initial time $t_i$. 
If $t_i=0$ then the initial density is too low for any amount of
smearing, including zero.  But with an appropriate choice of $t_i$, we
can reach the scaling solution rather quickly.  In every case the
scaling solution is reached by the end of the simulation; the final
curves differ from each other by roughly the statistical errorbars
(not shown for clarity), and we have checked that the string density
does not change significantly if we continue the simulation to a
larger time.

\bibliographystyle{unsrt}
\bibliography{paper_refs}

\begin{thebibliography}{10}

\bibitem{Peccei:1977hh}
R.D. Peccei and Helen~R. Quinn.
\newblock {CP Conservation in the Presence of Instantons}.
\newblock {\em Phys.Rev.Lett.}, 38:1440--1443, 1977.

\bibitem{Peccei:1977ur}
R.D. Peccei and Helen~R. Quinn.
\newblock {Constraints Imposed by CP Conservation in the Presence of
  Instantons}.
\newblock {\em Phys.Rev.}, D16:1791--1797, 1977.

\bibitem{Weinberg:1977ma}
Steven Weinberg.
\newblock {A New Light Boson?}
\newblock {\em Phys.Rev.Lett.}, 40:223--226, 1978.

\bibitem{Wilczek:1977pj}
Frank Wilczek.
\newblock {Problem of Strong p and t Invariance in the Presence of Instantons}.
\newblock {\em Phys.Rev.Lett.}, 40:279--282, 1978.

\bibitem{tHooft:1976}
Gerard 't~Hooft.
\newblock {Computation of the Quantum Effects Due to a Four-Dimensional
  Pseudoparticle}.
\newblock {\em Phys.Rev.}, D14:3432--3450, 1976.

\bibitem{Jackiw:1976pf}
R.~Jackiw and C.~Rebbi.
\newblock {Vacuum Periodicity in a Yang-Mills Quantum Theory}.
\newblock {\em Phys. Rev. Lett.}, 37:172--175, 1976.

\bibitem{Callan:1979bg}
Curtis~G. Callan, Jr., Roger~F. Dashen, and David~J. Gross.
\newblock {Instantons as a Bridge Between Weak and Strong Coupling in {QCD}}.
\newblock {\em Phys. Rev.}, D20:3279, 1979.

\bibitem{Preskill:1982cy}
John Preskill, Mark~B. Wise, and Frank Wilczek.
\newblock {Cosmology of the Invisible Axion}.
\newblock {\em Phys. Lett.}, B120:127--132, 1983.

\bibitem{Abbott:1982af}
L.~F. Abbott and P.~Sikivie.
\newblock {A Cosmological Bound on the Invisible Axion}.
\newblock {\em Phys. Lett.}, B120:133--136, 1983.

\bibitem{Dine:1982ah}
Michael Dine and Willy Fischler.
\newblock {The Not So Harmless Axion}.
\newblock {\em Phys. Lett.}, B120:137--141, 1983.

\bibitem{Davis:1986xc}
Richard~Lynn Davis.
\newblock {Cosmic Axions from Cosmic Strings}.
\newblock {\em Phys. Lett.}, B180:225, 1986.

\bibitem{Sikivie:1983ip}
P.~Sikivie.
\newblock {Experimental Tests of the Invisible Axion}.
\newblock {\em Phys.Rev.Lett.}, 51:1415--1417, 1983.

\bibitem{Bradley:2003kg}
R.~Bradley, J.~Clarke, D.~Kinion, L.~J. Rosenberg, K.~van Bibber, S.~Matsuki,
  M.~Muck, and P.~Sikivie.
\newblock {Microwave cavity searches for dark-matter axions}.
\newblock {\em Rev. Mod. Phys.}, 75:777--817, 2003.

\bibitem{Asztalos:2009yp}
S.J. Asztalos et~al.
\newblock {A SQUID-based microwave cavity search for dark-matter axions}.
\newblock {\em Phys.Rev.Lett.}, 104:041301, 2010.

\bibitem{Borsanyi:2015cka}
S.~Borsanyi, M.~Dierigl, Z.~Fodor, S.~D. Katz, S.~W. Mages, D.~Nogradi,
  J.~Redondo, A.~Ringwald, and K.~K. Szabo.
\newblock {Axion cosmology, lattice QCD and the dilute instanton gas}.
\newblock {\em Phys. Lett.}, B752:175--181, 2016.

\bibitem{diCortona:2015ldu}
Giovanni Grilli~di Cortona, Edward Hardy, Javier~Pardo Vega, and Giovanni
  Villadoro.
\newblock {The QCD axion, precisely}.
\newblock {\em JHEP}, 01:034, 2016.

\bibitem{Kibble}
T.~W.~B. Kibble.
\newblock {Topology of Cosmic Domains and Strings}.
\newblock {\em J. Phys.}, A9:1387--1398, 1976.

\bibitem{Albrecht:1984xv}
Andreas Albrecht and N.~Turok.
\newblock {Evolution of Cosmic Strings}.
\newblock {\em Phys. Rev. Lett.}, 54:1868--1871, 1985.

\bibitem{Dabholkar:1989ju}
Atish Dabholkar and Jean~M. Quashnock.
\newblock {Pinning Down the Axion}.
\newblock {\em Nucl. Phys.}, B333:815, 1990.

\bibitem{axion1}
Leesa Fleury and Guy~D. Moore.
\newblock {Axion dark matter: strings and their cores}.
\newblock {\em Journal of Cosmology and Astroparticle Physics}, 2016(01):004,
  2016.

\bibitem{Yamaguchi:1998gx}
Masahide Yamaguchi, M.~Kawasaki, and Jun'ichi Yokoyama.
\newblock {Evolution of axionic strings and spectrum of axions radiated from
  them}.
\newblock {\em Phys. Rev. Lett.}, 82:4578--4581, 1999.

\bibitem{Yamaguchi:1999yp}
Masahide Yamaguchi.
\newblock {Scaling property of the global string in the radiation dominated
  universe}.
\newblock {\em Phys. Rev.}, D60:103511, 1999.

\bibitem{Hiramatsu:2010yu}
Takashi Hiramatsu, Masahiro Kawasaki, Toyokazu Sekiguchi, Masahide Yamaguchi,
  and Jun'ichi Yokoyama.
\newblock {Improved estimation of radiated axions from cosmological axionic
  strings}.
\newblock {\em Phys.Rev.}, D83:123531, 2011.

\bibitem{Hiramatsu:2012gg}
Takashi Hiramatsu, Masahiro Kawasaki, Ken'ichi Saikawa, and Toyokazu Sekiguchi.
\newblock {Production of dark matter axions from collapse of string-wall
  systems}.
\newblock {\em Phys.Rev.}, D85:105020, 2012.

\bibitem{Vilenkin:1982ks}
A.~Vilenkin and A.~E. Everett.
\newblock {Cosmic Strings and Domain Walls in Models with Goldstone and
  PseudoGoldstone Bosons}.
\newblock {\em Phys. Rev. Lett.}, 48:1867--1870, 1982.

\bibitem{Martins:2000cs}
C.~J. A.~P. Martins and E.~P.~S. Shellard.
\newblock {Extending the velocity dependent one scale string evolution model}.
\newblock {\em Phys. Rev.}, D65:043514, 2002.

\bibitem{Hecht:1990mv}
Matthew~W. Hecht and Thomas~A. DeGrand.
\newblock Radiation patterns from vortex-antivortex annihilation.
\newblock {\em Phys. Rev.}, D42:519--528, 1990.

\bibitem{PIC}
R.W. Hockney and J.W. Eastwood.
\newblock {\em {Computer Simulation Using Particles}}.
\newblock Taylor \& Francis, 1988.

\bibitem{Wantz:2009mi}
Olivier Wantz and E.~P.~S. Shellard.
\newblock {The Topological susceptibility from grand canonical simulations in
  the interacting instanton liquid model: Chiral phase transition and axion
  mass}.
\newblock {\em Nucl. Phys.}, B829:110--160, 2010.

\bibitem{Albrecht:1989mk}
Andreas Albrecht and Neil Turok.
\newblock {Evolution of Cosmic String Networks}.
\newblock {\em Phys. Rev.}, D40:973--1001, 1989.

\bibitem{Bennett:1989yp}
David~P. Bennett and Francois~R. Bouchet.
\newblock {High resolution simulations of cosmic string evolution. 1. Network
  evolution}.
\newblock {\em Phys. Rev.}, D41:2408, 1990.

\bibitem{Allen:1990tv}
Bruce Allen and E.~P.~S. Shellard.
\newblock {Cosmic string evolution: a numerical simulation}.
\newblock {\em Phys. Rev. Lett.}, 64:119--122, 1990.

\bibitem{Vanchurin:2005yb}
Vitaly Vanchurin, Ken Olum, and Alexander Vilenkin.
\newblock {Cosmic string scaling in flat space}.
\newblock {\em Phys. Rev.}, D72:063514, 2005.

\bibitem{Olum:2006ix}
Ken~D. Olum and Vitaly Vanchurin.
\newblock {Cosmic string loops in the expanding Universe}.
\newblock {\em Phys. Rev.}, D75:063521, 2007.

\bibitem{Kalb:1974yc}
Michael Kalb and Pierre Ramond.
\newblock {Classical direct interstring action}.
\newblock {\em Phys. Rev.}, D9:2273--2284, 1974.

\bibitem{Vilenkin:1986ku}
Alexander Vilenkin and Tanmay Vachaspati.
\newblock {Radiation of Goldstone Bosons From Cosmic Strings}.
\newblock {\em Phys. Rev.}, D35:1138, 1987.

\bibitem{Turok:1984db}
Neil Turok and Pijushpani Bhattacharjee.
\newblock {Stretching Cosmic Strings}.
\newblock {\em Phys. Rev.}, D29:1557, 1984.

\bibitem{Shellard:1987bv}
E.~P.~S. Shellard.
\newblock {Cosmic String Interactions}.
\newblock {\em Nucl. Phys.}, B283:624--656, 1987.

\end{thebibliography}

\end{document}